\documentclass[12pt]{article}

  \usepackage{graphicx}
  \usepackage{epsfig}
  \usepackage{amssymb}
  \usepackage{subfigure}
  \usepackage{psfrag}
\evensidemargin - 1.truecm
\begin{document}

\newcommand{\be}{\begin{equation}}
\newcommand{\ee}{\end{equation}}
\newcommand{\nn}{\nonumber}
\newcommand{\bea}{\begin{eqnarray}}
\newcommand{\eea}{\end{eqnarray}}
\newcommand{\bfig}{\begin{figure}}
\newcommand{\efig}{\end{figure}}
\newcommand{\bc}{\begin{center}}
\newcommand{\ec}{\end{center}}
\newcommand{\bd}{\begin{displaymath}}
\newcommand{\ed}{\end{displaymath}}

\long\def\symbolfootnote[#1]#2{\begingroup%
\def\thefootnote{\fnsymbol{footnote}}\footnote[#1]{#2}\endgroup}

\begin{titlepage}
\nopagebreak
{\flushright{
        \begin{minipage}{5cm}
        Freiburg-THEP 05/06\\
        {\tt hep-ph/0507047}\\
        \end{minipage}        }

}                      
\vskip 3.5cm
\begin{center}
\boldmath
{\Large \bf Two-Loop Bhabha Scattering in QED
 }\unboldmath
\vskip 1.5cm
{\large  R.~Bonciani\symbolfootnote[3]{Email: 
{\tt Roberto.Bonciani@physik.uni-freiburg.de}}} and
{\large A.~Ferroglia\symbolfootnote[4]{Email: 
{\tt Andrea.Ferroglia@physik.uni-freiburg.de}}}
\vskip .7cm
{\it Fakult\"at f\"ur Mathematik und Physik, 
Albert-Ludwigs-Universit\"at
Freiburg, \\ D-79104 Freiburg, Germany} 
\end{center}
\vskip 1.2cm


\begin{abstract}  
In the context of pure QED, we obtain analytic expressions for the 
contributions to the Bhabha scattering differential cross section 
at order $\alpha^4$ which originate from the interference of two-loop 
photonic vertices with tree-level diagrams and from the interference 
of one-loop photonic diagrams amongst themselves. The ultraviolet 
renormalization is carried out. The IR-divergent soft-photon emission 
corrections are evaluated and added to the virtual cross section. 
The cross section obtained in this manner is valid for on-shell electrons and 
positrons of finite mass, and for arbitrary values of the center of 
mass energy and momentum transfer. We provide the expansion of our 
results in powers of the electron mass, and we compare them with the 
corresponding expansion of the complete order $\alpha^4$ photonic 
cross section, recently obtained in \cite{Penin:2005kf}. As a by-product, 
we obtain the contribution to the Bhabha scattering differential 
cross section of the interference of the two-loop photonic  
boxes with the tree-level diagrams, up to terms suppressed by positive 
powers of the electron mass. We evaluate numerically the various 
contributions to the cross section, paying particular attention to the 
comparison between exact and expanded results.

\vskip .3cm 
\flushright{
        \begin{minipage}{12.8cm}
\hspace*{0.6cm} {\it PACS}: 11.15.Bt, 12.20.Ds
        \end{minipage}        }
\end{abstract}
\vfill
\end{titlepage}

\section{Introduction \label{Intro}}

The Bhabha scattering process plays a crucial role in the study of elementary
particle phenomenology, since it is the process employed in the luminosity
measurement at $e^{+} e^{-}$ colliders. The small-angle Bhabha scattering at 
high ($\sim 100$ GeV) center of mass energy and the large-angle Bhabha 
scattering at intermediate ($1-10$ GeV) energies have cross sections that are 
large and QED dominated; these two characteristics allow for precise 
experimental measurements, as well as for a detailed theoretical evaluation of 
the cross section. 

The radiative corrections to the Bhabha scattering in pure QED have been 
extensively studied (see \cite{reviews} and references therein). The 
$\mathcal{O}(\alpha^3)$ corrections have been known for a long time, 
even in the full Electroweak Standard Model \cite{Bhabha1loop}. 
The second order QED corrections, $\mathcal{O}(\alpha^4)$, 
were the subject of renewed interest in the last few years, and several
works have been devoted to the study of second order 
radiative corrections, both virtual and real, enhanced by factors of 
$\ln^n(s/m^2)$ (with $n=1,2$, $s$ the c. m. energy, and $m$ the mass of
the electron) \cite{russians,Fadin:1993ha,Arbuzov:1995vj}.  The complete set 
of these corrections was finally obtained in \cite{Bas} by employing the 
QED virtual corrections for massless electron and positrons of \cite{Bern}, 
the results of \cite{bmr}, and by using the known 
structure of the IR poles in dimensional regularization \cite{Catani}. Very
recently, the complete set of photonic  $\mathcal{O}(\alpha^4)$ corrections 
to the cross section that are not suppressed by positive powers of the ratio 
$m^2/s$ were obtained in \cite{Penin:2005kf}.
The subset of virtual corrections of $\mathcal{O}(\alpha^4)$ involving a closed
fermion loop, together with the corresponding soft-photon emission corrections,
were obtained in an analytic and non approximated form in \cite{us,us2,us3}. 
The results contained in these papers do not rely on any mass expansion, and
they are valid for arbitrary values of the c.~m. energy $s$, momentum
transfer $t$, and of the electron mass $m$. In \cite{us,us2,us3}, the
calculation of the relevant loop diagrams was performed by employing the
Laporta-Remiddi algorithm \cite{Lap} which takes advantage of the 
integration by parts 
\cite{IBP} and Lorentz-invariance \cite{LI} identities in order to 
reduce the problem 
to the calculation of a small set of master integrals. The master integrals
are calculated using the differential equations method \cite{DiffEq}; their 
expression is given in terms of harmonic polylogarithms \cite{HPLs}. 
Both IR and UV divergencies were regularized in the dimensional regularization 
scheme \cite{DimReg} and they appear, in the intermediate results, as
singularities in 
$(D-4)$, where $D$ is the dimension of the space-time. It is natural to  
apply the same approach to the calculation of the complete set of 
$\mathcal{O}(\alpha^4)$ virtual corrections. At present, the list of the master 
integrals required to complete the calculation is available in
\cite{Czakon:2004tg,Czakon:2004wm}. However, only very few of the master 
integrals related to the two-loop photonic box diagrams have thus far 
been calculated \cite{Gundrum,Czakon:2004tg,Czakon:2004wm}. 
The subset of second order radiative corrections due to the interference of
one-loop diagrams was studied in \cite{Fleischer:2002wa}.

In using the results of \cite{RoPieRem1,RoPieRem2}, 
we calculate in this paper the following ${\mathcal O}(\alpha^4)$ corrections 
to the differential
cross section in QED:
\begin{itemize}
\item corrections due to the interference of the two-loop photonic vertex
diagrams with the tree-level amplitude;
\item corrections due to the interference of one-loop photonic diagrams amongst
themselves;
\item corrections due to the emission of two real soft photons from a 
tree-level diagram and of one soft photon from a one-loop photonic diagram.
\end{itemize}
All of the contributions mentioned above are calculated for finite electron mass 
and for arbitrary values of the c.~m. energy $s$  and momentum transfer $t$. 
We employ dimensional regularization in order to regularize both UV and IR 
divergencies. The UV renormalization is carried out in the on-shell scheme.
By following the same technique of \cite{us3}, we pair virtual and soft-photon 
emission corrections in order to check the cancellation of the IR singularities.

We also expand our results, that retain the full dependence on the electron mass, 
in the limit in which the electron mass is negligible with respect to the
Mandelstam invariants. In this way, it is possible:
\begin{itemize}
\item  to prove that our results reproduce the correct small-angle Bhabha 
scattering cross section at ${\mathcal O}(\alpha^4)$, which is determined by 
the Dirac vertex form factor only \cite{Fadin:1993ha}; 
\item  to provide strong cross-checks of a large part of the result of
\cite{Penin:2005kf}; 
\item  to find, by subtracting our result from the cross section of 
\cite{Penin:2005kf}, the contribution to the cross section of the interference 
between two-loop photonic boxes and the tree-level amplitude;
\item  to investigate the numerical relevance of the terms suppressed by positive
powers of the electron mass.
\end{itemize}

This paper is structured as follows:
after a brief summary of our notation in Section~\ref{sec0}, in Section~\ref{sec1},
we discuss the irreducible two-loop vertex photonic corrections, providing an 
expression for their contribution to the virtual differential cross section. 
In Sections~\ref{sec2}, \ref{sec3}, and \ref{secN}, we calculate 
the interference of the reducible vertex diagrams with the tree-level amplitude, 
the interference of the one-loop vertex diagrams with themselves and with the 
one-loop box diagrams, respectively, and we obtain the corresponding contributions
to the virtual differential cross section. 
In Section~\ref{sec4}, we complete the analysis of 
the interference among one-loop diagrams by considering the interference of
one-loop boxes. In Section~\ref{IR}, we discuss the soft-photon emission at 
${\mathcal O}(\alpha^4)$ and, in Section~\ref{IRcancel}, we explicitly show how 
the cancellation of the IR divergencies works between virtual 
and soft corrections. 
In Section~\ref{LL}, we analyze the expansion of our results in the limit
$m^2/s \to 0$; we discuss the behavior of the cross section 
at small angle, compare our calculations 
with the results present in the literature, and discuss the numerical 
accuracy of the expansion. Section~\ref{summ} contains our conclusions.
In Appendix~\ref{AppA}, we collect the definition of some function introduced 
in the paper. Finally, in Appendix~\ref{AppB}, we provide the expressions of the
contribution of the two-loop photonic boxes  to the differential 
cross section at order $\alpha^4$ 
in the limit $m^2/s \to 0$; all of the 
functions of the Mandelstam variables
introduced and employed throughout the paper are available in electronic format
in \cite{file}.

\section{Kinematic, Notation and Conventions \label{sec0}}

In this paper, we employ the notation and conventions adopted in
\cite{us,us2,us3}, which are summarized in the present section.
We consider the  Bhabha scattering process:
\be
e^-(p_1) + e^+(p_2) \longrightarrow e^-(p_3) + e^+(p_4) \, ,
\ee
where $p_1$, $p_2$, $p_3$, and $p_4$ are the momenta of the incoming electron,
incoming positron, outgoing electron, and outgoing positron, respectively. All
of the external particles are on their mass-shell, i.~e. $p_i^2 = -m^2$,
($i=1,\ldots,4$), where $m$ is the electron mass.

The Mandelstam invariants $s$, $t$, and $u$ are related to the beam energy
($E$) and scattering angle in the center of mass frame of reference ($\theta$)
by the relations
\bea
s &\equiv& - P^2 \equiv -(p_1 + p_2)^2 = 4 E^2 \, , \\
t &\equiv& - Q^2 \equiv -(p_1 - p_3)^2 = -4 \left(E^2-m^2\right)
\sin^2{\frac{\theta}{2}} \, , \\
u &\equiv& - V^2 \equiv -(p_1 - p_4)^2 = -4 \left(E^2-m^2\right)
\cos^2{\frac{\theta}{2}} \, . 
\eea  
Moreover, the Mandelstam invariants satisfy the relation $s + t + u = 4 m^2$.

The analytic expressions for the the vertex and box form factors that we employ
in the rest of the paper are calculated in the non physical kinematic region
$s<0$, and are then analytically continued to the physical region $s > 4 m^2$. 
In order to express the results (for $s <0$) 
in a compact form, it is
convenient to introduce the dimensionless variables $x$, $y$, and $z$, defined
through the following relations:
\bea
s &\equiv& -m^2 \frac{(1-x)^2}{x} \, , \quad x = \frac{\sqrt{4 m^2 - s} -
\sqrt{-s}}{\sqrt{4 m^2 - s} + \sqrt{-s}} \, , \quad 0 \le x\le 1 \,, \\
t &\equiv& -m^2 \frac{(1-y)^2}{y} \, , \quad y = \frac{\sqrt{4 m^2 - t} -
\sqrt{-t}}{\sqrt{4 m^2 - t} + \sqrt{-t}} \, , \quad 0 \le y\le 1 \,,\\
u &\equiv& -m^2 \frac{(1-z)^2}{z} \, , \quad z = \frac{\sqrt{4 m^2 - u} -
\sqrt{-u}}{\sqrt{4 m^2 - u} + \sqrt{-u}} \,, \quad 0 \le z\le 1 \,. 
\eea
The analytic continuation of all the form factors to the physical region 
$s > 4 m^2$ can be readily obtained by replacing $x \rightarrow -x' + i
\epsilon$, where $\epsilon$ is an infinitesimal positive quantity and where
\be
x' = - \frac{\sqrt{s-4 m^2} - \sqrt{s}}{\sqrt{s-4 m^2} + \sqrt{s}} \, ,
\ee
(see \cite{us2} for a more detailed discussion).

The Bhabha scattering differential cross section, calculated by summing 
over the spins of the
final state and averaged over the spins of the initial one, can be expanded in
powers of the fine structure constant $\alpha$ as follows:
\be
\frac{d \sigma (s,t,m^2)}{d \Omega} = 
\frac{d \sigma_0 (s,t,m^2)}{d \Omega} + 
\sum_{i = V,S} \left[ \left(\frac{\alpha}{\pi}\right)
\frac{d \sigma_1^{(i)} (s,t,m^2)}{d \Omega} + \left(\frac{\alpha}{\pi}\right)^2
\frac{d \sigma_2^{(i)}(s,t,m^2)}{d \Omega} \right]\, ,
\ee
where the superscripts $V$ and $S$ indicate virtual and soft-photon emission
contributions, respectively, while the subscripts $0$, $1$, and $2$ label the 
tree-level, ${\mathcal O}(\alpha^3)$ and ${\mathcal O}(\alpha^4)$  corrections,
respectively.

The tree-level (${\mathcal O}(\alpha^2)$) and ${\mathcal O}(\alpha^3)$ corrections 
are well known (their explicit expressions are collected, for example, 
in \cite{us2,us3}). The contribution of ${\mathcal O}(\alpha^4)$ to the virtual 
cross section can be further split as follows:
\be
\!
\frac{d \sigma_2^{(V)}(s,t,m^2)}{d \Omega} \! = \!
\frac{d \sigma_2^{(V,{\tt ph \,  Boxes})}(s,t,m^2)}{d \Omega}  +  
\frac{d \sigma_2^{(V,{\tt ph \,  Vertices})}(s,t,m^2)}{d \Omega}  
+  \frac{d \sigma_2^{(V,N_F=1)}(s,t,m^2)}{d \Omega}, \label{2Lcs}
\ee
where the superscript $N_F=1$ indicates  the UV-renormalized diagrams 
including a closed fermion loop (calculated, together with the corresponding 
soft radiation diagrams, in \cite{us2,us3}). The superscript 
``${\tt ph \, \, \,Vertices}$'' indicates the contribution of all of the 
UV-renormalized photonic corrections which include at least one vertex diagram, 
while the superscript  ``${\tt ph \, \, \,Boxes}$'' indicates the contribution 
to the cross sections of the photonic corrections which include box diagrams 
only.

This paper is dedicated to the calculation of $d \sigma_2^{(V,{\tt ph \, \, \,
Vertices})}/d \Omega$, as well as to the calculation  
of the contribution to $d \sigma_2^{(V,{\tt ph \, \, \,
Boxes})}/d \Omega$ deriving from the interference of one-loop box diagrams 
amongst
themselves; the soft-photon emission corrections that cancel the residual IR 
poles in the quantities mentioned above is also discussed. By employing these
results and the findings of \cite{Penin:2005kf}, we also obtain  the contribution 
of the interference of the two-loop photonic boxes to the cross section at order
$\alpha^4$, up to terms of order $m^2/s$ excluded.

\section{Irreducible Two-Loop Vertex Corrections \label{sec1}}

\begin{figure}
\bc
\epsfig{file=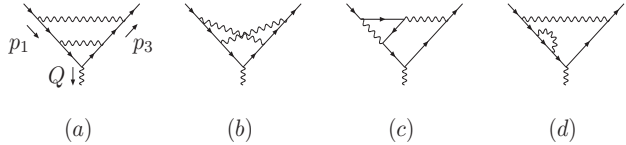,height=2.cm,width=4.5cm,
        bbllx=220pt,bblly=700pt,bburx=375pt,bbury=763pt}
\caption{\it Two-loop irreducible diagrams contributing to the photonic vertex
corrections of the electron current in the $t$-channel. The two diagrams,
analogous to Figs~(c) and~(d), which have an insertion of a one-loop vertex and
self-energy correction on the outgoing electron line, are not shown.}
\label{fig1}
\ec
\end{figure}

In this section, we obtain the contribution of the irreducible photonic vertex
corrections to the Bhabha scattering differential cross section at order 
$\alpha^4$. Four of the two-loop irreducible vertex graphs, which correct the 
electron current in the t-channel photon exchange contribution to the Bhabha
scattering, are shown in Fig.~\ref{fig1}. 
%
There are two other  graphs contributing to the process: these are the specular 
image of diagrams \ref{fig1}-(c) and \ref{fig1}-(d) with the fermionic arrow 
reversed, and their contribution to the differential cross section is 
identical to that of the diagrams in \ref{fig1}-(c) and  \ref{fig1}-(d). 
The sum of the graphs \ref{fig1}-(a) and \ref{fig1}-(b), and of the graphs 
\ref{fig1}-(c) and \ref{fig1}-(d) are separately  gauge independent.

The explicit expressions of diagrams \ref{fig1}-(a)--\ref{fig1}-(d) were
calculated in \cite{RoPieRem2}. The electron current which includes the 
two-loop photonic corrections can be written as
\be
\Gamma^\mu(p_1,p_3) = \left(\frac{\alpha}{\pi}\right)^2 \left[
F_1^{(2l,{\it ph})}(t) \gamma^\mu + \frac{1}{2 m} 
F_2^{(2l,{\it ph})}(t) \sigma^{\mu \nu} \left(p_1 -p_3\right) 
\right] \, , 
\label{ecurrent}
\ee
where $\sigma^{\mu \nu} = -i/2 [\gamma^\mu,\gamma^\nu]$. The electron spinors
and the dependence of the form factors $F_i^ {(2l,{\it ph})}(t)$ ($i=1,2$) on the
electron mass $m$ are omitted in Eq.~(\ref{ecurrent}).
The expression of the contribution of the single diagrams to the UV-renormalized
form factors $F_i^ {(2l,{\it ph})}(t)$ can be found in \cite{RoPieRem2}, and in
\cite{file}. The form factors shown in Eq.~(\ref{ecurrent}) still include 
IR singularities, which are regularized within the dimensional regularization 
scheme. The Laurent expansion of the form factors is
\bea
F_1^{(2l,{\it ph})}(t) &=& \frac{F_1^{(2l,{\it ph},-2)}(t)  }{(D-4)^2} + 
\frac{F_1^{(2l,{\it ph},-1)}(t) }{(D-4)}  + 
F_1^{(2l,{\it ph},0)}(t)  + {\mathcal O}(D-4)\, , \nn \\
 F_2^{(2l,{\it ph})}(t) &=& \frac{F_2^{(2l,{\it ph},-1)}(t)  }{(D-4)} + 
 F_2^{(2l,{\it ph},0)}(t) + {\mathcal O}(D-4)\, , \label{lauF}
\eea
where $D$ is the dimensional regulator.

\begin{figure}
\bc
\epsfig{file=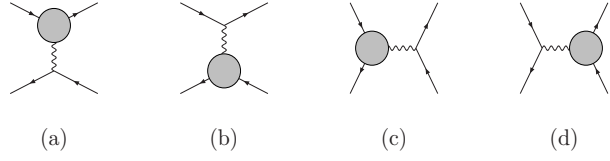,height=2.5cm,width=4.5cm,
        bbllx=220pt,bblly=680pt,bburx=375pt,bbury=763pt}
\caption{\it Two-loop photonic vertex diagrams contributing to  
Bhabha scattering.}
\label{fig2}
\ec
\end{figure}

The four  photonic two-loop vertex correction diagrams contributing to Bhabha 
scattering are shown in Fig.~\ref{fig2}, where the shaded circle represents the 
sum of the vertex graphs shown\footnote{The 
diagrams in Fig.~\ref{fig1}-(c) and \ref{fig1}-(d) enter the sum with a multiplicity  
factor of $2$.} in Fig.~\ref{fig1}.

The interference of diagrams \ref{fig2}-(b) and \ref{fig2}-(d) with the tree-level 
amplitude provides a contribution to the differential cross section
that is  identical to the one of diagrams \ref{fig2}-(a) and \ref{fig2}-(c),
respectively. This can easily be  proved by observing that diagram \ref{fig2}-(b) 
(\ref{fig2}-(d)) can be obtained from \ref{fig2}-(a) (\ref{fig2}-(c)) by
applying the transformations $p_2 \leftrightarrow -p_3$ and $p_4 \leftrightarrow
-p_1$, and that these transformations leave the Mandelstam invariants $s$ and
$t$ unchanged.

The contribution of the diagram in Fig.~\ref{fig2}-(a) to the Bhabha scattering
differential cross section at order $\alpha^4$ can be written as
\be
\frac{d \sigma_2^{(V,{\tt ph \,\,\,Vertices})}}{d \Omega} \Big|_{({\tt ph\,
Irr.\,Ver.},2-a)} =
\frac{\alpha^2}{s} \left[\frac{1}{s t} V_1^{(2l,{\tt Irr.\, Ver.})}(s,t) + 
\frac{1}{t^2} V_2^{(2l,{\tt Irr.\,Ver.})}(s,t) \right] \, .
\ee
The functions $V_i^{(2l,{\tt Irr. \, Ver.})}(s,t)$ ($i=1,2$) 
have the Laurent expansion
\bea
V_i^{(2l,{\tt Irr. \, Ver.})}(s,t) &=& \frac{V_i^{(2l,{\tt Irr. \,
Ver.},-2)}(s,t)}{(D-4)^2} + \frac{V_i^{(2l,{\tt Irr. \,
Ver.},-1)}(s,t)}{(D-4)}   \nn \\ 
& & + V_i^{(2l,{\tt Irr. \,
Ver.},0)}(s,t) + {\mathcal O}(D-4) \, ,
\eea 
with
\bea
V_i^{(2l,{\tt Irr. \,Ver.},-2)}(s,t) &=& c_{i1}(s,t) 
\mbox{Re} F_1^{(2l,{\it ph},-2)}(t) \, ,
\label{VphVa}\\ 
V_i^{(2l,{\tt Irr. \,Ver.},-1)}(s,t) &=& c_{i1}(s,t) 
\mbox{Re} F_1^{(2l,{\it ph},-1)}(t) +c_{i2}(s,t) 
\mbox{Re} F_1^{(2l,{\it ph},-2)}(t) \nn \\ 
& & + c_{i3}(s,t)  
\mbox{Re} F_2^{(2l,{\it ph},-1)}(t) \, ,
\label{VphVb}\\
V_i^{(2l,{\tt Irr. \,Ver.},0)}(s,t) &=& c_{i1}(s,t) 
\mbox{Re} F_1^{(2l,{\it ph},0)}(t)  + c_{i2}(s,t) 
\mbox{Re} F_1^{(2l,{\it ph},-1)}(t) \nn \\ 
& & + c_{i4}(s,t) 
\mbox{Re} F_1^{(2l,{\it ph},-2)}(t) + c_{i3}(s,t) 
\mbox{Re} F_2^{(2l,{\it ph}, 0)}(t) \nn \\ 
& & + c_{i6}(s,t) 
\mbox{Re} F_2^{(2l,{\it ph}, -1)}(t) \, .
\label{VphV}
\eea
The functions $c_{i j}$ are polynomials of the Mandelstam variables;
their explicit expressions are collected in Appendix~\ref{AppA}, while 
the Laurent coefficients of the form factors  $F_i^{(2l,{\it ph})}$ 
($i=1,2$) were introduced in Eq.~(\ref{lauF}). 
Even if the form factors are real for a physical (space-like) $t$, we write
$\mbox{Re} F_i^{(2l,{\it ph})}(t) $ ($i=1,2$) in the equations above
for convenience of later use.

The contribution to the differential cross section of the interference of 
the diagram in Fig.~\ref{fig2}-(c) with the tree-level amplitude can be 
obtained from the contribution of diagram \ref{fig2}-(a) by replacing $p_2
\leftrightarrow -p_3$. This is equivalent to exchange $s \leftrightarrow t$,
so that one finds
\bea
\frac{d \sigma_2^{(V,{\tt ph \,\,\,Vertices})}}{d \Omega} \Big|_{({\tt ph\,
Irr.\,Ver.},2-c)} =
\frac{\alpha^2}{s} \left[\frac{1}{s^2} V_2^{(2l,{\tt Irr.\, Ver.})}(t,s) + 
\frac{1}{s t} V_1^{(2l,{\tt Irr.\,Ver.})}(t,s) \right] \, . \label{V2c}
\eea
According to the definitions in Eqs.~(\ref{VphVa},\ref{VphVb},\ref{VphV}), 
the r.~h.~s. of Eq.~(\ref{V2c}) involves the functions 
$F_i^{(2l,{\it ph})}(s)$, which develop then an imaginary part above threshold 
($s>4 m^2$). These imaginary parts do not contribute to the differential 
cross section at this order.

Finally, the total contribution of the four diagrams in Fig.~\ref{fig2} to the
Bhabha scattering differential cross section at order $\alpha^4$ is
\bea
\frac{d \sigma_2^{(V,{\tt ph \,\,\,Vertices})}}{d \Omega} \Big|_{({\tt ph\,
Irr.\,Ver.})} &=& 
2 \frac{\alpha^2}{s} \left[\frac{1}{s^2} V_2^{(2l,{\tt Irr.\, Ver.})}(t,s) +
\frac{1}{t^2} V_2^{(2l,{\tt Irr.\, Ver.})}(s,t)  \right. \nn \\ 
& & + \left. \frac{1}{s t} \left( V_1^{(2l,{\tt Irr.\,Ver.})}(s,t) +
V_1^{(2l,{\tt Irr.\,Ver.})}(t,s)  \right)\right] \, .  \label{IrrVert}
\eea
The first, second, and third term within squared brackets in the r.~h.~s. of
Eq.~(\ref{IrrVert}) are the $s$-$s$, $t$-$t$, and $s$-$t$ channel interference 
amplitudes, respectively. 


\section{Reducible Two-Loop Vertex Corrections \label{sec2}}

\begin{figure}
\bc
\epsfig{file=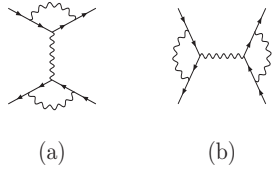,height=2.cm,width=4.5cm,
        bbllx=220pt,bblly=700pt,bburx=375pt,bbury=763pt}
\vspace*{.6cm}
\caption{\it Two-loop reducible vertex diagrams contributing to  
Bhabha scattering.}
\label{fig3}
\ec
\end{figure}

In this section, we consider the interference of the two-loop reducible diagrams,
shown in Fig.~\ref{fig3}, with the tree-level Bhabha scattering amplitude.
%
%
The two reducible diagrams contain a one-loop vertex correction in both
fermionic currents. Their contribution to the differential cross section at order
$\alpha^4$ can be written in terms of the one-loop UV-renormalized vertex form 
factors $F_i^{(1l)}$ ($i=1,2$) (see \cite{us2,file}) 
and of the functions $c_{ij}$ ($i=1,2$,$j=1,6$)
introduced in the previous section. It must also  
be observed that  the term
linear in $(D-4)$ is needed in the Laurent expansion 
of the one-loop vertex form factors. 

The contribution of diagram \ref{fig3}-(a) to the cross section is given by
\be
\frac{d \sigma_2^{(V,{\tt ph \,\,\,Vertices})}}{d \Omega} \Big|_{({\tt ph\,
Red.\,Ver.,\ref{fig3}-(a)})} \! \! = 
\frac{\alpha^2}{s} \left[\frac{1}{s t} V_1^{(2l,{\tt Red.\, Ver.})}(s,t) + 
\frac{1}{t^2} V_2^{(2l,{\tt Red.\,Ver.})}(s,t) \right] \, , 
\ee
where the Laurent expansion of $V_i^{(2l,{\tt Red.\, Ver.})}$ is 
\be
V_i^{(2l,{\tt Red.\, Ver.})} \! = \! 
\frac{V_i^{(2l,{\tt Red.\, Ver.},-2)}}{(D-4)^2} \! + \! 
\frac{V_i^{(2l,{\tt Red.\, Ver.},-1)}}{(D-4)} \! 
+ \!  V_i^{(2l,{\tt Red.\, Ver.},0)} + {\mathcal O}(D-4) , 
\ee
with
\bea
V_i^{(2l,{\tt Red.\, Ver.},-2)}(s,t)\!\! &=&\!\!   c_{i1}(s,t) \left[
\left(\mbox{Re} F_1^{(1l,-1)}(t) \right)^2 - \left(\mbox{Im} F_1^{(1l,-1)}(t) \right)^2 
\right] \, ,\\
V_i^{(2l,{\tt Red.\, Ver.},\!-\!1)}(s,t)\!\! &=& \!\! c_{i2}(s,t) \left[
\left(\mbox{Re} F_1^{(1l,\!-\!1)}(t) \right)^2 \!-\! \left(\mbox{Im} F_1^{(1l,\!-\!1)}(t) \right)^2
\right] \nn \\ 
& & \!\! + 2 c_{i1}(s,t)\!\left[\! \mbox{Re} F_1^{(1l,\!-\!1)}(t) \mbox{Re} F_1^{(1l,0)}(t) \!-\! 
\mbox{Im} F_1^{(1l,\!-\!1)}(t) \mbox{Im} F_1^{(1l,0)}(t)\!\right] \nn \\
& &\!\! + 2 c_{i3}(s,t)\!\left[\! \mbox{Re} F_1^{(1l,\!-\!1)}(t) \mbox{Re} F_2^{(1l,0)}(t) \!-\! 
\mbox{Im} F_1^{(1l,\!-\!1)}(t) \mbox{Im} F_2^{(1l,0)}(t)\!\right] \, ,
\nn \\
& & \\
V_i^{(2l,{\tt Red.\, Ver.},0)}(s,t)\!\! &=&  c_{i4}(s,t) \left[
\left(\mbox{Re} F_1^{(1l,\!-\!1)}(t) \right)^2 \!-\! \left(\mbox{Im} F_1^{(1l,\!-\!1)}(t) 
\right)^2 \right] \nn \\ 
& &\!\! + 2 c_{i2}(s,t)\!\left[\! \mbox{Re} F_1^{(1l,\!-\!1)}(t) \mbox{Re} F_1^{(1l,0)}(t) \!-\! 
\mbox{Im} F_1^{(1l,\!-\!1)}(t) \mbox{Im} F_1^{(1l,0)}(t)\!\right] \nn \\
& &\!\! + 2 c_{i6}(s,t)\!\left[\! \mbox{Re} F_1^{(1l,\!-\!1)}(t) \mbox{Re} F_2^{(1l,0)}(t) \!-\! 
\mbox{Im} F_1^{(1l,\!-\!1)}(t) \mbox{Im} F_2^{(1l,0)}(t)\!\right] \nn \\
& &\!\! + 2 c_{i1}(s,t)\!\left[\! \mbox{Re} F_1^{(1l,\!-\!1)}(t) \mbox{Re} F_1^{(1l,1)}(t) \!-\! 
\mbox{Im} F_1^{(1l,\!-\!1)}(t) \mbox{Im} F_1^{(1l,1)}(t)\!\right] \nn \\
& &\!\! + 2 c_{i3}(s,t)\!\left[\! \mbox{Re} F_1^{(1l,\!-\!1)}(t) \mbox{Re} F_2^{(1l,1)}(t) \!-\! 
\mbox{Im} F_1^{(1l,\!-\!1)}(t) \mbox{Im} F_2^{(1l,1)}(t)\!\right] \nn \\
& &\!\! + c_{i1}(s,t) \left[
\left(\mbox{Re} F_1^{(1l,0)}(t) \right)^2 \!-\! \left(\mbox{Im} F_1^{(1l,0)}(t) 
\right)^2 \right] \nn \\ 
& &\!\! + 2 c_{i3}(s,t)\!\left[\! \mbox{Re} F_1^{(1l,0)}(t) \mbox{Re} F_2^{(1l,0)}(t) \!-\! 
\mbox{Im} F_1^{(1l,0)}(t) \mbox{Im} F_2^{(1l,0)}(t)\!\right] \nn \\
& &\!\! + c_{i5}(s,t) \left[
\left(\mbox{Re} F_2^{(1l,0)}(t) \right)^2 \!-\! \left(\mbox{Im} F_2^{(1l,0)}(t) 
\right)^2 \right] 
\, .
\eea

Similarly, the contribution of diagram \ref{fig3}-(b) to the cross section at
order $\alpha^4$ is given by
\be
\frac{d \sigma_2^{(V,{\tt ph \,\,\,Vertices})}}{d \Omega} \Big|_{({\tt ph\,
Red.\,Ver.,\ref{fig3}-(b)})} = 
\frac{\alpha^2}{s} \left[\frac{1}{s t} V_1^{(2l,{\tt Red.\, Ver.})}(t,s) + 
\frac{1}{s^2} V_2^{(2l,{\tt Red.\,Ver.})}(t,s) \right] \, .
\ee

It is then possible to conclude that the  interference of the diagrams in 
Fig.~\ref{fig3} with the tree-level amplitude gives the following contribution to
the cross sections
\bea
\frac{d \sigma_2^{(V,{\tt ph \,\,\,Vertices})}}{d \Omega} \Big|_{({\tt ph\,
Red.\,Ver.)}} &=& 
\frac{\alpha^2}{s} \left[\frac{1}{s^2} V_2^{(2l,{\tt Red.\, Ver.})}(t,s) +
\frac{1}{t^2} V_2^{(2l,{\tt Red.\, Ver.})}(s,t)  \right. \nn \\ 
& & + \left. \frac{1}{s t} \left( V_1^{(2l,{\tt Red.\,Ver.})}(s,t) +
V_1^{(2l,{\tt Red.\,Ver.})}(t,s)  \right)\right] \, . 
\label{RedVertBB}
\eea
The first, second, and third term within squared brackets in the r.~h.~s. of
Eq.~(\ref{RedVertBB}) are the $s$-$s$, $t$-$t$, and $s$-$t$ channel interference 
amplitudes, respectively.


\section{Interference of One-Loop Vertex Diagrams \label{sec3}}
\begin{figure}
\bc
\epsfig{file=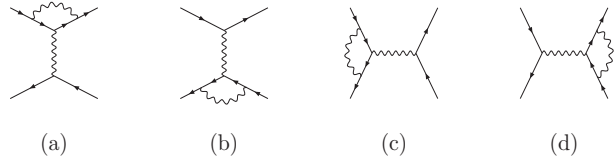,height=2.5cm,width=4.5cm,
        bbllx=220pt,bblly=680pt,bburx=375pt,bbury=763pt}
\caption{\it One-loop vertex diagrams contributing to  
Bhabha scattering.}
\label{fig4}
\ec
\end{figure}

In this section, we obtain the Bhabha scattering cross section at order 
$\alpha^4$ deriving from the interference of the diagrams in Fig.~\ref{fig4}.

We begin by considering the contribution of the amplitude of diagram
\ref{fig4}-(a) squared; one finds that
\be
\frac{d \sigma_2^{(V,{\tt ph \,\,\,Vertices})}}{d \Omega} \Big|_{({\tt ph\,
Ver. Ver.,\ref{fig4}-(a)})} = 
\frac{\alpha^2}{s} \frac{1}{t^2} I_{1}(s,t) \, . \label{LLa}
\ee 
In Eq.~(\ref{LLa}), we introduce the function $I_1$ that has the following 
Laurent expansion in powers of $(D-4)$:
\bea
I_1(s,t) &=& \frac{I_1^{(-2)}(s,t)}{(D-4)^2} + \frac{I_1^{(-1)}(s,t)}{(D-4)} + 
 I_1^{(0)}(s,t) + {\mathcal O}(D-4) \, ,
\eea
with
\bea
I_1^{(-2)}(s,t) &=& \frac{1}{2} c_{21}(s,t)\left[
\left(\mbox{Re} F_1^{(1l,-1)}(t) \right)^2 + \left(\mbox{Im} F_1^{(1l,-1)}(t) \right)^2 
\right] \, , \\
I_1^{(-1)}(s,t) &=&\frac{1}{2} c_{22}(s,t)\left[
\left(\mbox{Re} F_1^{(1l,-1)}(t) \right)^2 + \left(\mbox{Im} F_1^{(1l,-1)}(t) \right)^2 
\right]\nn \\
& &\!\! + c_{21}(s,t)\!\left[\! \mbox{Re} F_1^{(1l,-1)}(t) \mbox{Re}
F_1^{(1l,0)}(t) + 
\mbox{Im} F_1^{(1l,-1)}(t) \mbox{Im} F_1^{(1l,0)}(t)\!\right]\nn \\
& & \!\! + c_{23}(s,t)\!\left[\! \mbox{Re} F_1^{(1l,-1)}(t) \mbox{Re}
F_2^{(1l,0)}(t) + 
\mbox{Im} F_1^{(1l,-1)}(t) \mbox{Im} F_2^{(1l,0)}(t)\!\right]
\, ,\\
I_1^{(0)}(s,t) &=& \!\! c_{22}(s,t)\!\left[\! \mbox{Re} F_1^{(1l,-1)}(t) \mbox{Re}
F_1^{(1l,0)}(t) + 
\mbox{Im} F_1^{(1l,-1)}(t) \mbox{Im} F_1^{(1l,0)}(t)\!\right]\nn \\
& &\!\! + c_{21}(s,t)\!\left[\! \mbox{Re} F_1^{(1l,-1)}(t) \mbox{Re}
F_1^{(1l,1)}(t) + 
\mbox{Im} F_1^{(1l,-1)}(t) \mbox{Im} F_1^{(1l,1)}(t)\!\right]\nn \\
& &\!\! + c_{22}(s,t)\!\left[\! \mbox{Re} F_1^{(1l,-1)}(t) \mbox{Re}
F_2^{(1l,0)}(t) + 
\mbox{Im} F_1^{(1l,-1)}(t) \mbox{Im} F_2^{(1l,0)}(t)\!\right]\nn \\
& &\!\! + c_{23}(s,t)\!\left[\! \mbox{Re} F_1^{(1l,-1)}(t) \mbox{Re}
F_2^{(1l,1)}(t) + 
\mbox{Im} F_1^{(1l,-1)}(t) \mbox{Im} F_2^{(1l,1)}(t)\!\right]\nn \\
& &\!\! + \frac{1}{2} c_{21}(s,t)\left[
\left(\mbox{Re} F_1^{(1l,0)}(t) \right)^2 + \left(\mbox{Im} F_1^{(1l,0)}(t) \right)^2 
\right] \nn \\
& &\!\! + c_{23}(s,t)\!\left[\! \mbox{Re} F_1^{(1l,0)}(t) \mbox{Re}
F_2^{(1l,0)}(t) + 
\mbox{Im} F_1^{(1l,0)}(t) \mbox{Im} F_2^{(1l,0)}(t)\!\right] \nn \\
& &\!\! + c_{27}(s,t)\left[
\left(\mbox{Re} F_2^{(1l,0)}(t) \right)^2 + \left(\mbox{Im} F_2^{(1l,0)}(t) \right)^2 
\right] \, . 
\eea
%
%
The expression of the functions $c_{ij}$ in terms of the Mandelstam
invariants can be found in Appendix~\ref{AppA}, while the expression of
the one-loop vertex form factors appearing in the functions $I_1$ can be found
in \cite{us2,RoPieRem2} and are collected in \cite{file}.

Also the contribution of the square of the diagram in Fig.~\ref{fig4}-(c) can be
easily expressed by employing the function $I_1$:
\be
\frac{d \sigma_2^{(V,{\tt ph \,\,\,Vertices})}}{d \Omega} \Big|_{({\tt ph\,
Ver. Ver.,\ref{fig4}-(c)})} = 
\frac{\alpha^2}{s} \frac{1}{s^2} I_{1}(t,s) \, . \label{LLc}
\ee

The interference between the diagrams in Fig.~\ref{fig4}-(a) and 
Fig.~\ref{fig4}-(b) generates a term in the cross section that can be
written as
\be
\frac{d \sigma_2^{(V,{\tt ph \,\,\,Vertices})}}{d \Omega} \Big|_{({\tt ph\,
Ver. Ver.,\ref{fig4}-(a)-(b)})} = 
\frac{\alpha^2}{s} \frac{1}{t^2} I_{2}(s,t) \, , \label{LLab}
\ee
where the function $I_2$ is given by
\bea
I_2(s,t) &=& \frac{I_2^{(-2)}(s,t)}{(D-4)^2} + \frac{I_2^{(-1)}(s,t)}{(D-4)} + 
 I_2^{(0)}(s,t) + {\mathcal O}(D-4) \, , 
\eea
with
\bea
I_2^{(-2)}(s,t)  &=& 2 I_1^{(-2)}(s,t) \, , \\
I_2^{(-1)}(s,t)  &=& 2 I_1^{(-1)}(s,t) \, , \\
I_2^{(0)}(s,t)  &=& 2 I_1^{(0)}(s,t) +  
\left(c_{25}(s,t) - 2 c_{27}(s,t)\right) \times \nn \\ 
& & \times \left[
\left(\mbox{Re} F_2^{(1l,0)}(t) \right)^2 + \left(\mbox{Im} F_2^{(1l,0)}(t) \right)^2\, 
\right] \, .
\eea

Similarly, the interference between the diagrams in Fig.~\ref{fig4}-(c) and
Fig.~\ref{fig4}-(d) generates the following contribution to the differential
cross section
\be
\frac{d \sigma_2^{(V,{\tt ph \,\,\,Vertices})}}{d \Omega} \Big|_{({\tt ph\,
Ver. Ver.,\ref{fig4}-(c)-(d)})} = 
\frac{\alpha^2}{s} \frac{1}{s^2} I_{2}(t,s) \, .
\label{LLcd}
\ee

The interference of the diagrams in Fig.~\ref{fig4}-(a) and Fig.~\ref{fig4}-(c)
can be expressed in terms of a third function, $I_3$:
\be
\frac{d \sigma_2^{(V,{\tt ph \,\,\,Vertices})}}{d \Omega} \Big|_{({\tt ph\,
Ver. Ver.,\ref{fig4}-(a)-(c)})} = 
\frac{\alpha^2}{s} \frac{1}{s t} I_{3}(s,t) \, . \label{LLac}
\ee
Also in this case, it is convenient to explicitly write the Laurent expansion 
of the function $I_3$:
\bea
I_3(s,t) &=& \frac{I_3^{(-2)}(s,t)}{(D-4)^2} + \frac{I_3^{(-1)}(s,t)}{(D-4)} + 
 I_3^{(0)}(s,t) + {\mathcal O}(D-4) \, , 
\eea
with
\bea
I_3^{(-2)}(s,t) &=&  c_{11}(t,s) F_1^{(1l,-1)}(t) \mbox{Re} F_1^{(1l,-1)}(s)  \,
, \\
I_3^{(-1)}(s,t) &=& c_{12}(t,s) F_1^{(1l,-1)}(t) \mbox{Re} F_1^{(1l,-1)}(s)  \nn \\
& & + c_{11}(t,s) \left[F_1^{(1l,-1)}(t) \mbox{Re} F_1^{(1l,0)}(s)  +
 F_1^{(1l,0)}(t) \mbox{Re} F_1^{(1l,-1)}(s) \right] \nn \\
& & + c_{13}(t,s) \left[F_1^{(1l,-1)}(t) \mbox{Re} F_2^{(1l,0)}(s) +
F_2^{(1l,0)}(t) \mbox{Re} F_1^{(1l,-1)}(s)\right] \, ,\\
I_3^{(0)}(s,t) &=& c_{14}(t,s) F_1^{(1l,-1)}(t) \mbox{Re} F_1^{(1l,-1)}(s)  \nn \\
& & + c_{12}(t,s) \left[F_1^{(1l,-1)}(t) \mbox{Re} F_1^{(1l,0)}(s)  +
 F_1^{(1l,0)}(t) \mbox{Re} F_1^{(1l,-1)}(s) \right] \nn \\
& & + c_{11}(t,s) \left[F_1^{(1l,-1)}(t) \mbox{Re} F_1^{(1l,1)}(s)  +
 F_1^{(1l,1)}(t) \mbox{Re} F_1^{(1l,-1)}(s) \right] \nn \\
& & + c_{16}(t,s) F_1^{(1l,-1)}(t) \mbox{Re} F_2^{(1l,0)}(s)  +c_{16}(s,t)
 F_2^{(1l,0)}(t) \mbox{Re} F_1^{(1l,-1)}(s) \nn \\
& & + c_{13}(t,s) F_1^{(1l,-1)}(t) \mbox{Re} F_2^{(1l,1)}(s)  +
c_{13}(s,t)  F_2^{(1l,1)}(t) \mbox{Re} F_1^{(1l,-1)}(s)  \nn \\
& & + c_{13}(s,t) F_2^{(1l,0)}(t) \mbox{Re} F_1^{(1l,0)}(s)  +
c_{13}(t,s)  F_1^{(1l,0)}(t) \mbox{Re} F_2^{(1l,0)}(s)  \nn \\
& & + c_{17}(t,s) F_2^{(1l,0)}(t) \mbox{Re} F_2^{(1l,0)}(s) +
c_{11}(t,s)  F_1^{(1l,0)}(t) \mbox{Re} F_1^{(1l,0)}(s)\, .
\eea

Finally, it is easy to verify that all of the interferences between 
$s$- and $t$-channel diagrams in Fig.~\ref{fig4} give a contribution to the
cross section identical to the one in Eq.~(\ref{LLac}), while the squared
amplitude of the diagram \ref{fig4}-(b) (\ref{fig4}-(d)) coincides with the
squared amplitude of diagram \ref{fig4}-(a) (\ref{fig4}-(c)). We can then
conclude that the total contribution of the interferences of the diagrams in
Fig.~\ref{fig4} to the Bhabha scattering cross section is given by
\bea
\frac{d \sigma_2^{(V,{\tt ph \,\,\,Vertices})}}{d \Omega} \Big|_{({\tt ph\,
Ver. Ver.)}} &=& \frac{\alpha^2}{s} \left[ \frac{1}{t^2} \left( 2 I_1(s,t) +
I_2(s,t)\right) +  \frac{4}{s t} I_3 (s,t) \right. \nn \\ 
& & + \left.\frac{1}{s^2} \left( 2 I_1(t,s) + 
I_2(t,s)\right) \right] \, .
\eea 
%


\section{Interference of One-Loop Vertex and One-Loop Box Diagrams \label{secN}}

\begin{figure}
\bc
\epsfig{file=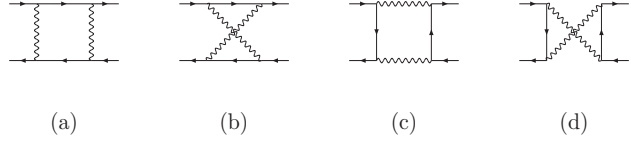,height=2.5cm,width=4.5cm,
        bbllx=220pt,bblly=680pt,bburx=375pt,bbury=763pt}
\caption{\it One-loop box diagrams contributing to  
Bhabha scattering.}
\label{fig5}
\ec
\end{figure}

In the present section, we consider the interference of the diagrams in
Fig.~\ref{fig4} with the diagrams in Fig.~\ref{fig5}.

In order to write down the contributions that these interferences provide to the 
Bhabha scattering cross sections at order $\alpha^4$, it is convenient to
introduce the functions $I_{4,i}(x_1,x_2,x_3)$, where $i=1,\ldots,3$ and where
$x_j$ ($j=1,\ldots,3$) represents one of the Mandelstam invariants $s$,$t$, and $u$; 
these functions have the following Laurent expansion in $(D-4)$:
\bea
I_{4,i}(x_1,x_2,x_3) \!=\! \frac{I_{4,i}^{(-2)}(x_1,x_2,x_3)}{(D-4)^2} \!+\! 
\frac{I_{4,i}^{(-1)}(x_1,x_2,x_3)}{(D-4)} + I_{4,i}^{(0)}(x_1,x_2,x_3) \!+\!
{\mathcal O}(D-4) \, .
\eea
The coefficients of the Laurent expansion are given by
\bea
I_{4,i}^{(-2)}(x_1,x_2,x_3) &=& x_3 \mbox{Re} \Biggl[ \left(F_1^{(1l,-1)}(x_1)\right)^* 
B_i^{(1l,-1)}(x_2,x_3) \Biggr] \, , 
 \\
I_{4,i}^{(-1)}(x_1,x_2,x_3) &=&  x_3\mbox{Re} 
\Biggl[\left(F_1^{(1l,-1)}(x_1)\right)^* B_i^{(1l,0)}(x_2,x_3) 
 + \left(F_1^{(1l,0)}(x_1)\right)^* B_i^{(1l,-1)}(x_2,x_3) \nn
\\
& & + \left(F_2^{(1l,0)}(x_1)\right)^* B_{i+3}^{(1l,-1)}(x_2,x_3)\Biggr]\, ,  \\
I_{4,i}^{(0)}(x_1,x_2,x_3) &=& x_3\mbox{Re}
 \Biggl[\left(F_1^{(1l,-1)}(x_1)\right)^* B_i^{(1l,1)}(x_2,x_3)
 + \left(F_1^{(1l,0)}(x_1)\right)^* B_i^{(1l,0)}(x_2,x_3) \nn \\
& & + \left(F_1^{(1l,1)}(x_1)\right)^* B_i^{(1l,-1)}(x_2,x_3) 
+ \left(F_2^{(1l,0)}(x_1)\right)^* B_{i+3}^{(1l,0)}(x_2,x_3) \nn \\
& & + \left(F_2^{(1l,1)}(x_1)\right)^* B_{i+3}^{(1l,-1)}(x_2,x_3)\Biggr]\,. 
\eea
%
%

Besides for the one-loop vertex form factors already employed 
in the previous sections, one 
encounters in the equations above the functions $B_i^{(1l,j)}$ 
($i=1,\ldots,3$, $j=-1,0,1$) introduced in \cite{us2} to describe the
contribution of the diagrams in Fig.~\ref{fig5} to the Bhabha scattering 
cross section\footnote{While the analytic expressions of the
functions $B_i^{(1l,-1)}$ and  $B_i^{(1l,0)}$ are explicitly given in 
\cite{us2} the expressions of the functions $B_i^{(1l,1)}$ are  not.
They are the coefficients of $(D-4)$ in the Laurent expansion of the functions 
$B_i^{(1l)}$ (Eq.~(45) in \cite{us2}), that were not needed in that context.
Their expression in terms of HPLs and dimensionless variables can be
found in \cite{file}.} at ${\mathcal O}(\alpha^3)$. The auxiliary functions 
$B_i^{(1l,j)}$ ($i=4,\ldots,6$, $j=-1,0$) are introduced for the first time here, and
their expressions in terms of HPLs and the dimensionless variables 
$x$, $y$, and $z$ can be found in \cite{file}.

The interference of diagram \ref{fig4}-(a) with diagram \ref{fig5}-(a) gives the
following contribution to the differential cross section:
\bea
\frac{d \sigma_2^{(V,{\tt ph \,\,\,Vertices})}}{d \Omega} \Big|_{({\tt ph\,
Ver. Box,\ref{fig4}-(a)-\ref{fig5}-(a)})} = 
\frac{\alpha^2}{4 s} \frac{1}{t^2}  I_{4,2}(t,s,t) \, . \label{BVaa}
\eea
The notation has been chosen to clarify that Eq.~(\ref{BVaa}) is related to the
interference of two $t$-channel diagrams.

Similarly, it is possible to write all the interferences of diagrams
\ref{fig4}-(a) and \ref{fig4}-(c) with the diagrams in Fig.~\ref{fig5} in terms
of the functions $I_{4,i}$:
\bea
\frac{d \sigma_2^{(V,{\tt ph \,\,\,Vertices})}}{d \Omega} \Big|_{({\tt ph\,
Ver. Box,\ref{fig4}-(a)-\ref{fig5}-(b)})} &=& 
-\frac{\alpha^2}{4 s} \frac{1}{t^2}I_{4,2}(t,u,t) \, , \label{BVab} \\
\frac{d \sigma_2^{(V,{\tt ph \,\,\,Vertices})}}{d \Omega} \Big|_{({\tt ph\,
Ver. Box,\ref{fig4}-(a)-\ref{fig5}-(c)})} &=& 
\frac{\alpha^2}{4 s} \frac{1}{s t} I_{4,1}(t,t,s) \, , \label{BVac} \\
\frac{d \sigma_2^{(V,{\tt ph \,\,\,Vertices})}}{d \Omega} \Big|_{({\tt ph\,
Ver. Box,\ref{fig4}-(a)-\ref{fig5}-(d)})} &=& 
\frac{\alpha^2}{4 s} \frac{1}{s t}I_{4,3}(t,u,s) \, , \label{BVad} \\
\frac{d \sigma_2^{(V,{\tt ph \,\,\,Vertices})}}{d \Omega} \Big|_{({\tt ph\,
Ver. Box,\ref{fig4}-(c)-\ref{fig5}-(a)})} &=& 
\frac{\alpha^2}{4 s} \frac{1}{s t} I_{4,1}(s,s,t) \, , \label{BVca} \\
\frac{d \sigma_2^{(V,{\tt ph \,\,\,Vertices})}}{d \Omega} \Big|_{({\tt ph\,
Ver. Box,\ref{fig4}-(c)-\ref{fig5}-(b)})} &=& 
\frac{\alpha^2}{4 s} \frac{1}{s t} I_{4,3}(s,u,t) \, , \label{BVcb} \\
\frac{d \sigma_2^{(V,{\tt ph \,\,\,Vertices})}}{d \Omega} \Big|_{({\tt ph\,
Ver. Box,\ref{fig4}-(c)-\ref{fig5}-(c)})} &=& 
\frac{\alpha^2}{4 s} \frac{1}{s^2}I_{4,2}(s,t,s) \, , \label{BVcc} \\
\frac{d \sigma_2^{(V,{\tt ph \,\,\,Vertices})}}{d \Omega} \Big|_{({\tt ph\,
Ver. Box,\ref{fig4}-(c)-\ref{fig5}-(c)})} &=& 
-\frac{\alpha^2}{4 s} \frac{1}{s^2} I_{4,2}(s,u,s) \, . \label{BVcd}
\eea

The interference of diagram \ref{fig4}-(b) (\ref{fig4}-(d))
with a diagram in Fig.~\ref{fig5} is identical to the interference of diagram 
\ref{fig4}-(a) (\ref{fig4}-(c)) with the same diagram in Fig.~\ref{fig5}.
Therefore, the total contribution of the interferences between one-loop vertex diagrams
and one-loop box diagrams to the Bhabha scattering cross section is twice the
sum of Eqs.~(\ref{BVaa}-\ref{BVcd}); namely
\bea
\frac{d \sigma_2^{(V,{\tt ph \,\,\,Vertices})}}{d \Omega} \Big|_{({\tt ph\,
Ver. Box)}} \!\!&=&\frac{\alpha^2}{2 s} \Biggl[ \frac{1}{t^2} 
 \left(I_{4,2}(t,s,t) - I_{4,2}(t,u,t)\right)\nn \\
 & & \!\!\!
- \frac{1}{s t} \left( I_{4,1}(t,t,s) +\!I_{4,3}(t,u,s) + \!I_{4,1}(s,s,t)+\!
I_{4,3}(s,u,t)\right)  \nn \\ 
& & \!\!\!
+ \frac{1}{s^2} \left( I_{4,2}(s,t,s)  - I_{4,2}(s,u,s) \right) \Biggr]\, .
\eea


\section{Interference of One-Loop Box Diagrams \label{sec4}}

With reference to Eq.~(\ref{2Lcs}), it is possible to further split 
$\sigma_2^{(V,{\tt ph \, Boxes})}$ in the sum of two contributions; the first originates from the
interference of two-loop photonic box diagrams and tree-level diagrams, 
the second originates from 
the interference of two one-loop box diagrams:
\be
\frac{d \sigma_2^{(V,{\tt ph\,\,\,Boxes})}}{d \Omega} = 
\left. \frac{d \sigma_2^{(V,{\tt ph\,\,\,Boxes})}}{d \Omega} \right|_{({\tt ph\,
Box\, Box})} + 
\left. \frac{d \sigma_2^{(V,{\tt ph\,\,\,Boxes})}}{d \Omega} 
\right|_{({\tt 2 L \,Box})} \, . \label{2Lboxes}
\ee
The two-loop photonic box diagrams in the $m \neq 0$ case are at the
moment still unknown, and the calculation of the second term in the r.~h.~s. of   
Eq.~(\ref{2Lboxes}) currently  remains as an open problem (see \cite{Czakon:2004tg, Czakon:2004wm}). 

On the contrary, the calculation of the first term in  Eq.~(\ref{2Lboxes}) is, in 
principle, straightforward. The interference of every pair of the one-loop box diagrams 
shown in Fig.~\ref{fig5} provides a contribution to the differential cross section of the
form
\be
\left. \frac{d \sigma_2^{(V,{\tt ph\,\,\,Boxes})}}{d \Omega} \right|_{({\tt ph\,
Box\,Box}, i j)} = \frac{\alpha^2}{4 s} C_{ij}(x,y,z) \, ,
\ee
where the indices $i,j$ run over the diagram labels ($i,j=a,b,c,d$) and where
we have introduced new functions $C_{ij}$. The explicit expression of the
latter, already continued to the physical region $s> 4 m^2$, 
is particularly long and can be found in  \cite{file}.

\boldmath
\section{Soft-Photon Emission at Order $\alpha^4$ \label{IR}}
\unboldmath

All of the two-loop photonic corrections discussed in the previous sections are 
UV renormalized, but they still include double and single poles in $(D-4)$. 
These singularities have an IR origin, and they can be eliminated by adding 
the contribution of the real soft-photon emission 
diagrams at order  $\alpha^4$ to the virtual cross section.

Before discussing the soft corrections to the Bhabha scattering differential
cross section of order $\alpha^4$,  the reader is reminded that the soft
corrections at order $\alpha^3$, discussed in detail in \cite{us3}, can be
written in the factorized form
\bea
\left(\frac{\alpha}{\pi}\right) \frac{d \sigma_1^S(s,t,m^2)}{d \Omega} = 
\left(\frac{\alpha}{\pi}\right) \frac{d \sigma_0^D(s,t,m^2)}{d \Omega}
S_{\mathrm{IR}} \, ,
\eea
where $\sigma_0^D(s,t,m^2)$ is the tree-level cross section obtained by
calculating the traces of Dirac matrices  in $D$ dimensions and where
$S_{\mathrm{IR}}$ is
defined as
\bea
S_{{\mathrm{IR}}} &\equiv& 4 \sum_{j=1}^4 J_{1j} \,, \quad J_{1j} = \epsilon_j \left(
p_1 \cdot p_j\right) I_{1j}  \, , 
\label{sIR}
\eea
with $\epsilon_1 = \epsilon_4 =1$ and $\epsilon_2 = \epsilon_3 = -1$, and
\be
I_{1j} = \frac{1}{\Gamma\left(3-\frac{D}{2}\right) \pi^{(D-4)/2}}
\frac{m^{(D-4)}}{4 \pi^2} \int^\omega \frac{d^{D-1} k}{k_0} \frac{1}{\left(p_1 \cdot
k\right)\left(p_j \cdot k\right)} \, . \label{Iij}
\ee
The integral in Eq.~(\ref{Iij}) can be found in \cite{GP} (see also Appendix~A of 
\cite{us3}); the integration over the momentum of the soft photon ($k$) is
restricted to the region $|\vec{k}| = k_0 < \omega$, where $\omega$ is the 
cut-off on the energy of the unobserved soft photon.
The expansion of $\sigma_0^D(s,t,m^2)$ in powers of $(D-4)$ is
\bea
\frac{d \sigma_0^D(s,t,m^2)}{d \Omega}  &=& 
\frac{d \sigma_0(s,t,m^2)}{d \Omega} + 
(D-4) \,\,  \frac{d \sigma^{(1)}_0(s,t,m^2)}{d \Omega} \nn \\ 
& & + (D-4)^2 \,\, \frac{d \sigma^{(2)}_0(s,t,m^2)}{d \Omega} 
+ {\mathcal O}\left((D-4)^3\right) \, , 
\label{laurT}
\eea
where $\sigma_0(s,t,m^2)$ is the well known tree-level cross section, and
\bea
\frac{d \sigma^{(1)}_0(s,t,m^2)}{d \Omega} &=&\! \!\!\frac{\alpha^2}{s} \left\{ 
\frac{1}{s^2} \left[ \frac{s^2}{4} \right] \!+\! 
\frac{1}{t^2} \left[ \frac{t^2}{4} \right] \!+\!  
\frac{1}{s t} \left[ \frac{1}{2} (s\!+\!t)^2 -\frac{1}{2} s t - m^2 (s\!+\!t)\right]
\right\} , \label{bhaD1} \\
\frac{d \sigma^{(2)}_0(s,t,m^2)}{d \Omega} &=&\! \!\! \frac{\alpha^2}{s}
\frac{1}{s t} \left[ - \frac{1}{4}  s t\right] \label{bhaD2} \, .
\eea
The contribution of the $s$- and $t$- channel diagrams, and of their interference
to $\sigma^{(1)}_0(s,t,m^2)$ and $\sigma^{(2)}_0(s,t,m^2)$, is evident in 
Eqs.~(\ref{bhaD1},\ref{bhaD2}).

\begin{figure}
\bc
\epsfig{file=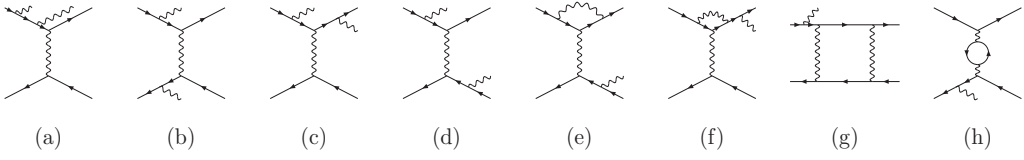,height=2.5cm,width=4.5cm,
        bbllx=220pt,bblly=680pt,bburx=375pt,bbury=763pt}
\caption{\it Examples of double photon emission from tree-level diagrams and
single photon emission from one-loop diagrams.}
\label{fig6}
\ec
\end{figure}

There are two different kinds of soft-photon emission diagrams contributing to the real
corrections to the cross section at  order $\alpha^4$: 
\begin{itemize}
\item[i)] the tree-level diagrams with the
emission of two soft photons (some diagrams belonging to this class are shown in
Fig.~\ref{fig6}-(a)--(d)), and 
\item[ii)] the diagrams which include a one-loop correction and the
emission of a soft photon from one of the external legs (see the examples 
in Fig.~\ref{fig6}-(e)--(h)). 
\end{itemize}
%
%

Since the soft-photon corrections in QED exponentiate, the contribution of the
double emission diagrams to the Bhabha scattering differential cross section is 
given by
\be
\left(\frac{\alpha}{\pi}\right)^2 
\frac{d \sigma_2^{(S,{\tt double})}(s,t,m^2)}{d \Omega} = 
\frac{1}{2} \left(\frac{\alpha}{\pi}\right)^2 \frac{d \sigma_0^D(s,t,m^2)}{d \Omega}
\left(S_{{\mathrm{IR}}} \right)^2 \, . 
\label{double}
\ee
The emission of a single photon from one-loop diagrams, interfered with
the tree-level single photon emission graphs, amounts to
\be
\left(\frac{\alpha}{\pi}\right)^2 
\frac{d \sigma_2^{(S,{\tt single})}(s,t,m^2)}{d \Omega} = 
\left(\frac{\alpha}{\pi}\right)^2 \frac{d \sigma_1^{(V,D)}(s,t,m^2)}{d
\Omega} 
S_{{\mathrm{IR}}}  \, , \label{a4single}
\ee
where $\sigma_1^{(V,D)}$ is the UV-renormalized virtual cross section at order
$\alpha^3$. The superscript $D$ was introduced as a reminder that 
the Laurent expansion of $\sigma_1^{(V,D)}$ must be known up to 
and including terms linear in
$(D-4)$; 
the reason for this is that these linear terms give rise to a
finite contribution when multiplied by the single pole present in $S_{{\mathrm{IR}}}$. 
The real corrections of order $\alpha^4$ which originate 
from single-photon emission
diagrams and which include a fermionic loop (as, for example, diagram \ref{fig6}-(h)),
were calculated  in \cite{us3}; they include single IR poles that cancel against 
the virtual corrections of order $\alpha^4$ that also include a photon self-energy 
insertion. This set does not play any role to the present discussion and is 
thus systematically ignored.

With the above in mind, the one-loop virtual cross section appearing in
Eq.~(\ref{a4single}) can be written as
\be
\frac{d \sigma_1^{(V,D)}(s,t,m^2)}{d \Omega}  = 
\frac{d \sigma_1^{(V,D)}(s,t,m^2)}{d \Omega} \Bigg|_{(1l,V)} + 
\frac{d \sigma_1^{(V,D)}(s,t,m^2)}{d \Omega} \Bigg|_{(1l,B)} \, , 
\label{1lV}
\ee
where the subscript $V$ indicates the contribution of the interference
between vertex graphs and tree-level amplitude; $B$ stands for 
the cross section generated by the interference between one-loop boxes
and tree-level diagrams. Furthermore, we split the terms on the r.~h.~s. of 
Eq.~(\ref{1lV}) as follows:
\bea
\frac{d \sigma_1^{(V,D)}(s,t,m^2)}{d \Omega} \Bigg|_{(1l,j)} &=&
\frac{d \sigma_1^{V}(s,t,m^2)}{d \Omega} \Bigg|_{(1l,j)} + (D-4)
\frac{d \sigma_1^{(V,1)}(s,t,m^2)}{d \Omega} \Bigg|_{(1l,j)} \nn \\ 
& & + {\mathcal O} \left( (D-4)^2 \right) \, , 
\eea
with $ j = V, B$. 
The singular and finite parts of the one-loop virtual cross section, 
corresponding to the first term in the r.~h.~s. of Eq.~(\ref{1lV}), are well
known. Their expression in terms of Mandelstam invariants and HPLs can
be found in Eq.~(43) and Eq.~(49) of \cite{us2}.
The terms proportional to $(D-4)$ which arise from box and vertex one-loop 
corrections are given by
\bea
\frac{d \sigma_1^{(V,1)}(s,t,m^2)}{d \Omega} \Bigg|_{(1l,B)} &=& 
\frac{\alpha^2}{s}\Biggl[ \frac{m^2}{s}\Bigl( \mbox{Re} B_1^{(1l,1)}(s,t)+
\mbox{Re} B_2^{(1l,1)}(t,s) + B_3^{(1l,1)}(u,t) \nn \\ 
& & - \mbox{Re} B_2^{(1l,1)}(u,s) \Bigr) + \frac{m^2}{t}\Bigl( \mbox{Re} B_2^{(1l,1)}(s,t)+
\mbox{Re} B_1^{(1l,1)}(t,s) \nn \\ 
& & - B_2^{(1l,1)}(u,t) +
 \mbox{Re}  B_3^{(1l,1)}(u,s) \Bigr) \Biggr] \, , \label{that}
\eea
and
\be
\frac{d \sigma_1^{(V,1)}(s,t,m^2)}{d \Omega} \Bigg|_{(1l,V)} \! \! \! = \! 
2 \frac{\alpha^2}{s} \Biggl[ \frac{1}{s^2}V_2^{(1l,1)}(t,s) \! + \! 
\frac{1}{t^2}V_2^{(1l,1)}(s,t) \! 
+ \! \frac{1}{s t} \left(V_1^{(1l,1)}(t,s) + V_1^{(1l,1)}(s,t)\right)
\Biggr] ,
\ee
respectively, with
\bea
V_i^{(1l,1)}(s,t) &=& c_{i1}(s,t) \mbox{Re} F_1^{(1l,1)} (t) +
c_{i2}(s,t) \mbox{Re} F_1^{(1l,0)} (t)
+ \, c_{i4}(s,t) \mbox{Re} F_1^{(1l,-1)} (t)  \nn \\
& & + c_{i6}(s,t) \mbox{Re} F_2^{(1l,0)}
(t) + \, c_{i3}(s,t) \mbox{Re} F_2^{(1l,1)} (t) \, , \quad\quad \quad (i =1,2) \, .
\label{this}
\eea
All of the functions appearing in the r.~h.~s. of Eqs.~(\ref{that}-\ref{this}) 
were introduced in the previous sections, and their expressions in terms of 
Mandelstam invariants and HPLs have been collected in \cite{file}.

As is mentioned above, the calculation of the integrals $I_{1j}$ which
appears in Eq.~(\ref{sIR}) has been carried out in \cite{us3}, up to terms
linear in $(D-4)$ excluded. At first glance, it appears that the 
calculation of such terms is needed, since they provide, in the limit $D
\rightarrow 4$, a non vanishing contribution to both Eq.~(\ref{double}) and
Eq.~(\ref{a4single}). However, it is possible to prove that  this is not the
case. In order to proceed with our proof,  one needs to split the
one-loop UV-renormalized  virtual corrections in an IR-divergent part and a
finite reminder, as is exemplified in the following:
\be
\frac{d \sigma_1^{V}(s,t,m^2)}{d \Omega} \Bigg|_{(1l,j)}  = 
\frac{1}{(D-4)} \frac{d \sigma_1^{(V,-1)}(s,t,m^2)}{d \Omega} \Bigg|_{(1l,j)} + 
\frac{d \sigma_1^{(V,0)}(s,t,m^2)}{d \Omega} \Bigg|_{(1l,j)} \,.
\label{laurV}
\ee
The Laurent expansion of $S_{\mathrm{IR}}$ has the form
\be
S_{\mathrm{IR}} = \frac{S^{(-1)}_{\mathrm{IR}}}{(D-4)} + 
S^{(0)}_{\mathrm{IR}} + 
(D-4) S^{(1)}_{\mathrm{IR}} + {\mathcal O}\left((D-4)^2\right)\, .
\label{laurrad}
\ee
The cancellation of IR divergencies in the order $\alpha^3$ cross section 
guarantees that\footnote{We remind the reader that the photon self-energy 
diagrams are IR-finite (see \cite{us3}).}
\be
\frac{d \sigma_1^{(V,-1)}(s,t,m^2)}{d \Omega} \Bigg|_{(1l,B)} + 
\frac{d \sigma_1^{(V,-1)}(s,t,m^2)}{d \Omega} \Bigg|_{(1l,V)} +
\frac{d \sigma_0(s,t,m^2)}{d \Omega} S^{(-1)}_{{\mathrm{IR}}} = 0\, .
\label{cancel}
\ee
By employing Eqs.~(\ref{double},\ref{a4single}) in combination with
Eqs.~(\ref{laurT},\ref{laurV},\ref{laurrad}), one can prove that the non-vanishing 
term proportional to $S^{(1)}_{{\mathrm{IR}}}$ appearing in the double
emission cross section (Eq.~(\ref{double})) is
\be
\frac{d \sigma_2^{(S,{\tt double})}(s,t,m^2)}{d \Omega} \rightarrow
\frac{d \sigma_0(s,t,m^2)}{d \Omega} S^{(-1)}_{{\mathrm{IR}}} 
S^{(1)}_{{\mathrm{IR}}} \, ,
\ee
while the non vanishing term proportional to $S^{(1)}_{{\mathrm{IR}}}$
appearing in the single-photon emission cross sections at order $\alpha^4$ 
is
\be
\frac{d \sigma_2^{(S,{\tt single})}(s,t,m^2)}{d \Omega} \rightarrow
\left[ 
\frac{d \sigma_1^{(V,-1)}(s,t,m^2)}{d \Omega} \Bigg|_{(1l,B)} + 
\frac{d \sigma_1^{(V,-1)}(s,t,m^2)}{d \Omega} \Bigg|_{(1l,V)}\right] 
S^{(1)}_{{\mathrm{IR}}} \, .
\ee
Therefore, we can conclude that, due to Eq.~(\ref{cancel}), the non 
vanishing terms proportional to $S^{(1)}_{{\mathrm{IR}}}$ cancel out
in the total real emission cross section at order
$\alpha^4$, given by the sum of Eqs.~(\ref{double},\ref{a4single}).


\section{Cancellation of the IR Singularities \label{IRcancel}}

The IR divergencies in the real corrections at order $\alpha^4$ 
(Eqs.~(\ref{double},\ref{a4single})) should cancel the IR singularities present
in the virtual corrections discussed in Sections~\ref{sec1}-\ref{sec4} and 
the ones arising from the interference of  the (yet unknown) two-loop photonic box 
graphs.

\begin{figure}
\bc
\hspace*{2mm}
\epsfig{file=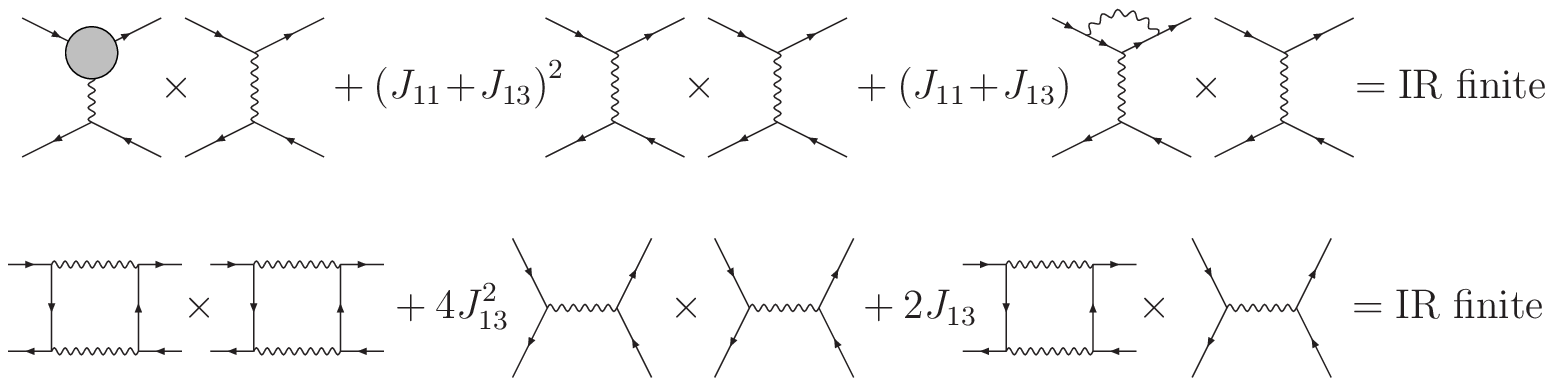,height=3.8cm,width=4.5cm,
        bbllx=220pt,bblly=632pt,bburx=375pt,bbury=763pt}
\caption{\it Example of the cancellation of the IR divergencies in the two-loop 
irreducible vertex corrections (first line) and in the interference of the
one-loop direct box in the $s$-channel with itself (second line).}
\label{figcanc}
\ec
\end{figure}
%
As in the case of the cancellation of the IR divergencies of the virtual
cross section at order $\alpha^4 (N_F=1)$ discussed in \cite{us3}, it is
possible to organize the contributions to the cross section in IR-finite blocks
by pairing the virtual corrections originating from a certain set of diagrams
with an appropriate subset of the soft-photon emission corrections.
As an
example, in the first line of Fig.~\ref{figcanc} we illustrate the cancellation of the IR poles 
present in the interference of the graph in Fig.~\ref{fig2}-(a) with the
$t$-channel tree-level diagram; in all the terms in the l.~h.~s., the product
of two graphs represents the  contribution of their interference to the Bhabha
scattering differential  cross section. Once again, the gray circle represents
the sum of the UV-renormalized  two-loop photonic vertex corrections to the
electron current in the  $t$-channel photon-exchange diagram. In the second
line of
Fig.~\ref{figcanc}, we provide another example of the cancellation of the IR
divergencies in the box-by-box sector.
Similar relations can be found for all of 
the contributions to the cross section at 
order $\alpha^4$.

\begin{figure}
\bc
\hspace*{-8mm}
\epsfig{file=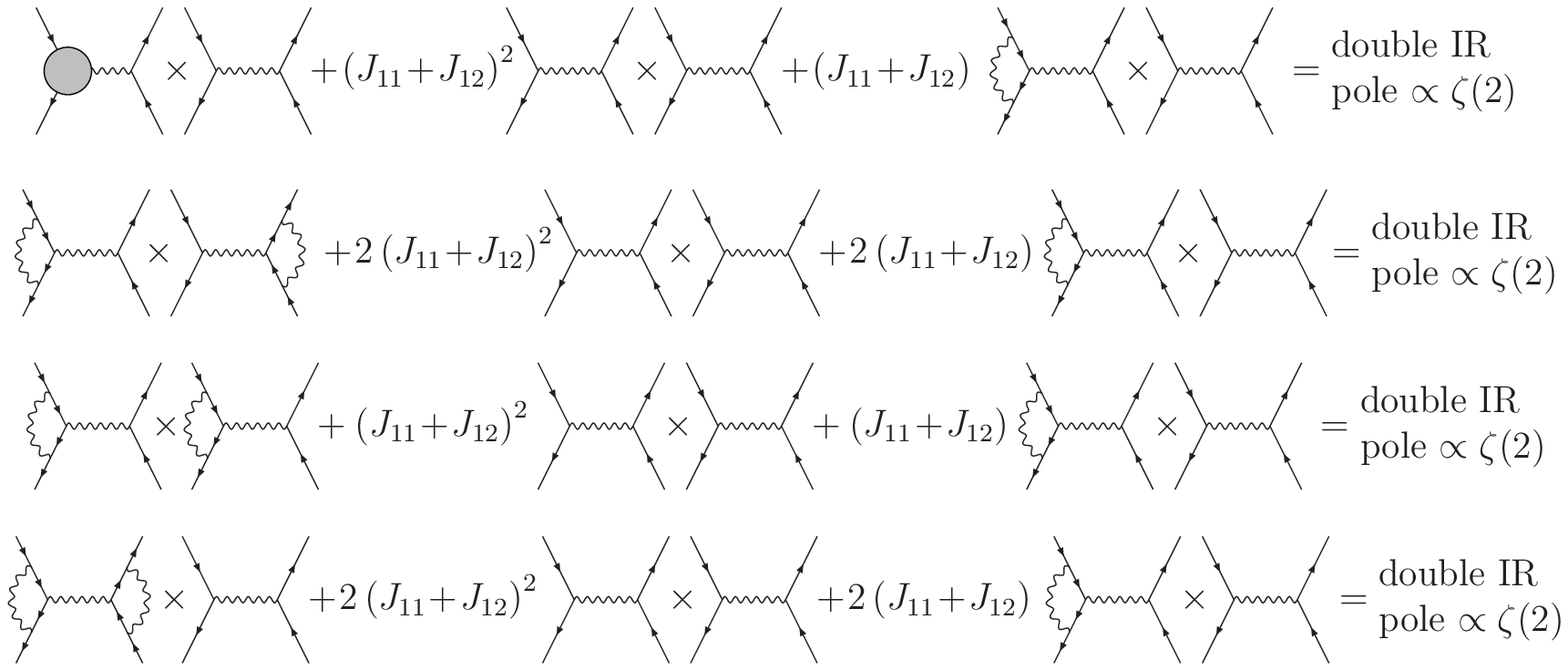,height=6.5cm,width=4.5cm,
        bbllx=220pt,bblly=530pt,bburx=375pt,bbury=763pt}
\caption{\it Residual IR poles proportional to $\zeta(2)$ in the $s$-channel
cross section.}
\label{figz2s2}
\ec
\end{figure}
%
\begin{figure}
\bc
\hspace*{-8mm}
\epsfig{file=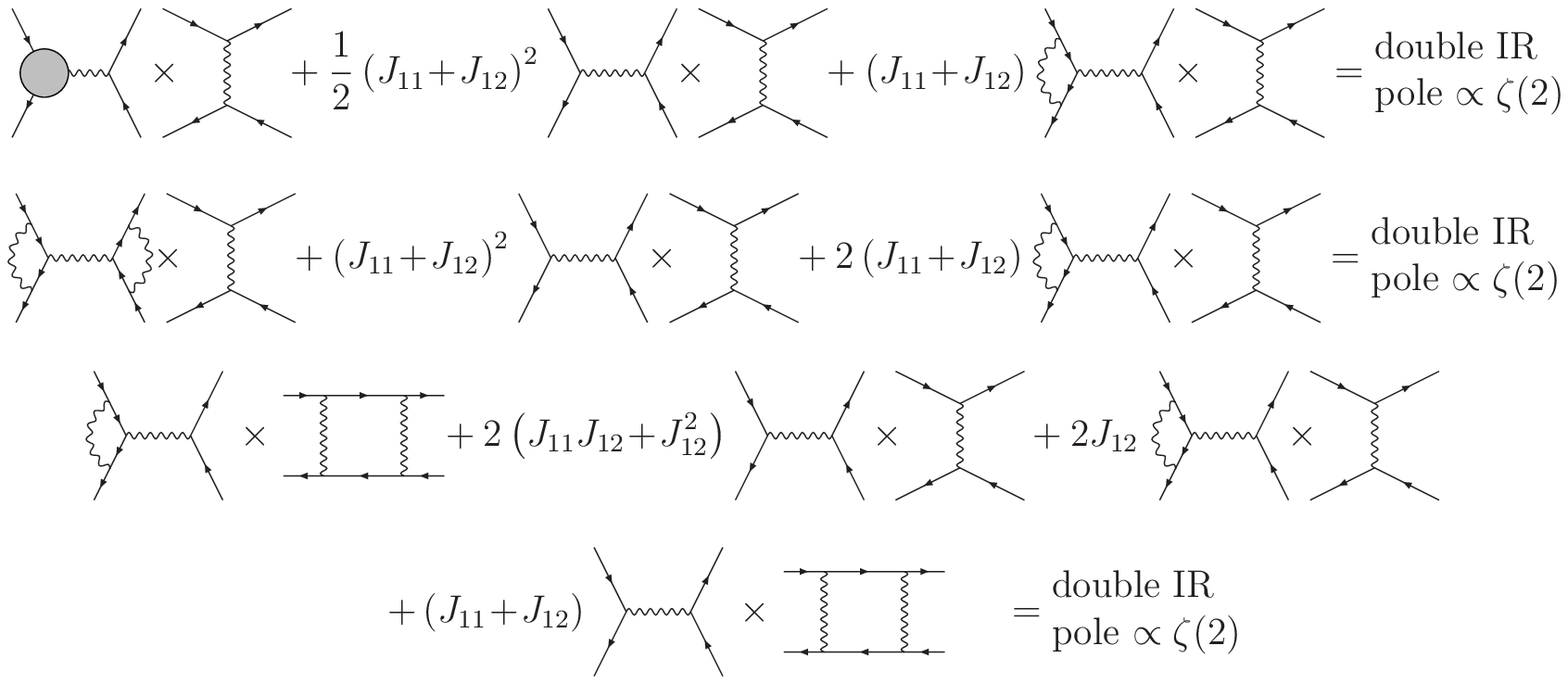,height=7.cm,width=4.5cm,
        bbllx=220pt,bblly=530pt,bburx=375pt,bbury=763pt}
\caption{\it Residual IR poles proportional to $\zeta(2)$ in the $s$-$t$-channel
interference contribution to $\sigma_{2}^{(V,{\tt ph \,\,\, Vertices})}$.}
\label{figz2st}
\ec
\end{figure}
%
A special case is represented by the virtual corrections in which the coefficients  
of the IR poles, when calculated in the non-physical region $s<0$, include HPLs with 
two or more zeroes in the rightmost positions in the weight list. When performing the
analytic continuation to the physical region $ s >  4 m^2$, these HPLs generate
real terms proportional to $\zeta(2)$, where $\zeta$ is the Riemann Zeta
function.  For example, by replacing the non physical dimensionless variable $x$
according to   $x \rightarrow -x' + i \epsilon$, one finds that
\be
H(0,0;x)  \rightarrow  H(0,0;-x' + i \epsilon) = 
H(0,0;x') - 3 \zeta(2)  + i \pi H(0;x') \, , 
\ee
or
\be
H(-1,0,0;x) \rightarrow H( -1,0,0;-x' + i \epsilon) =  -
H(1,0,0;x')+ 3 \zeta(2) - i \pi H(1,0;x') \, .
\ee
The part of the IR pole, that is proportional to the $\zeta(2)$ factor
arising from analytical continuation does not cancel in the combination with the
soft-radiation contributions that eliminate the other IR singularities. 
This kind
\begin{figure}
\bc
\epsfig{file=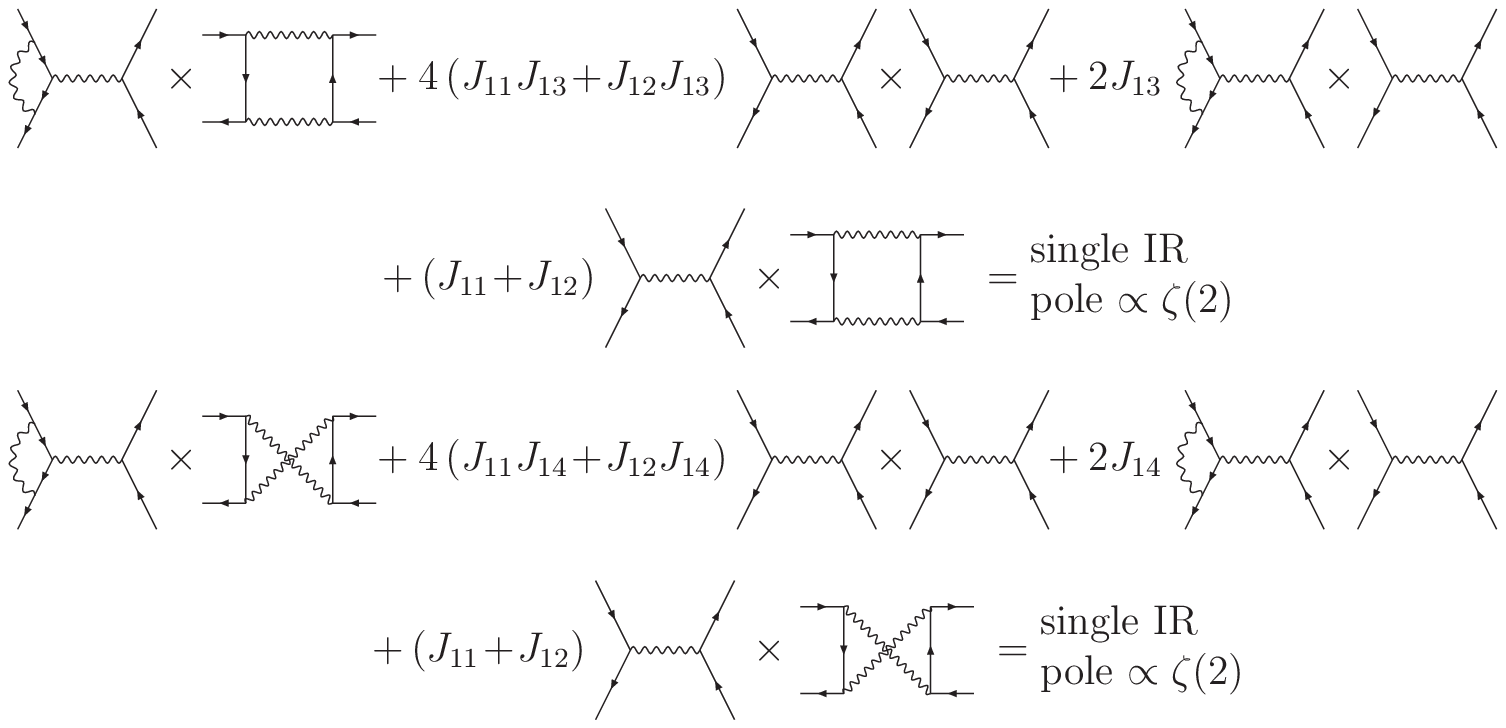,height=7.0cm,width=4.5cm,
        bbllx=220pt,bblly=530pt,bburx=375pt,bbury=763pt}
\caption{\it Residual IR single poles proportional to $\zeta(2)$ in the interference
of one-loop vertex by one-loop box diagrams.}
\label{figz23334}
\ec
\end{figure}
%
of behavior, already observed in \cite{us3}  in the discussion of the ${\mathcal
O}(\alpha^4 (N_F=1))$ cross section, is again encountered 
in the  cases illustrated in Figs.~\ref{figz2s2}--\ref{boxbyboxz2}.

%
%
\begin{figure}
\bc
\epsfig{file=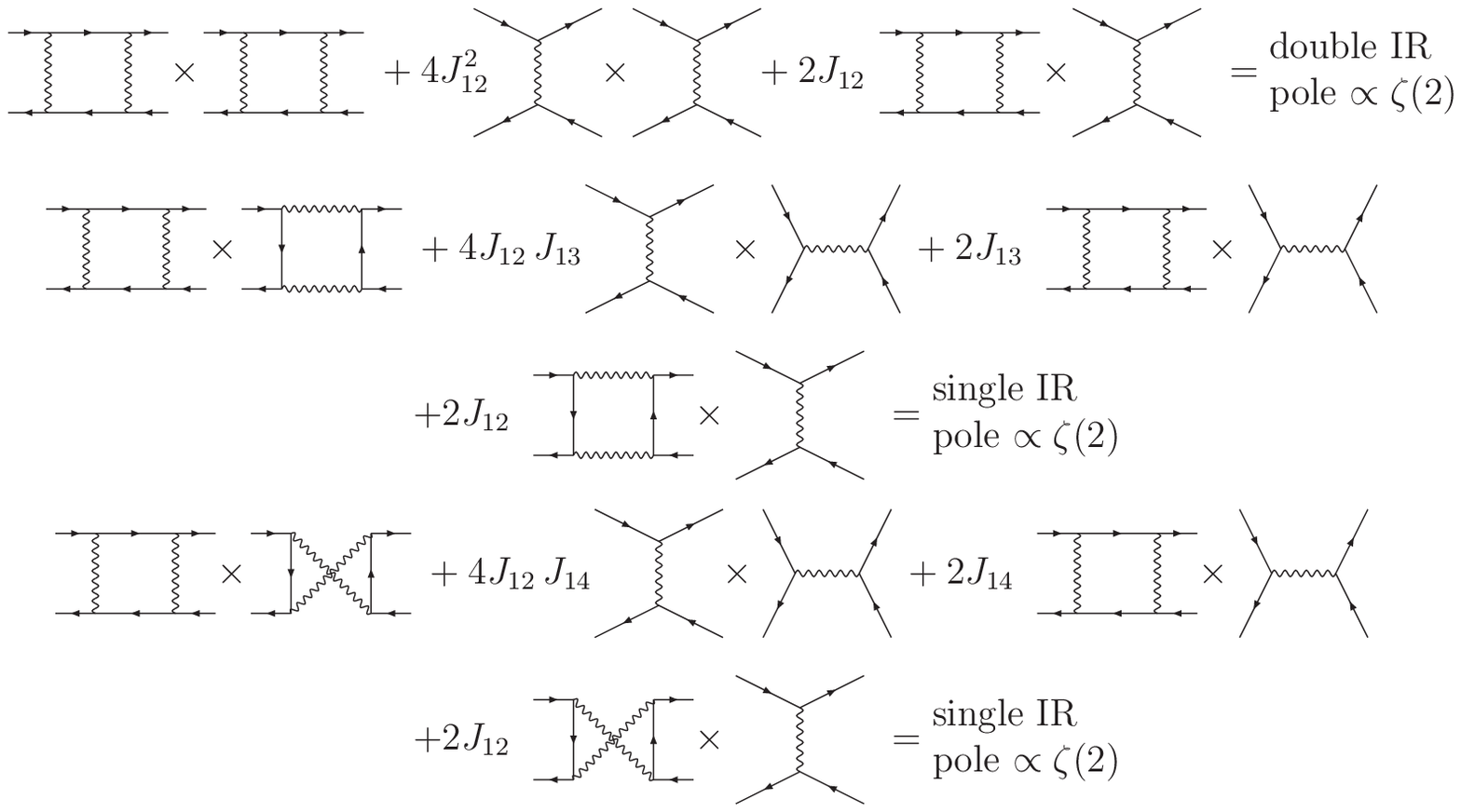,height=8.5cm,width=4.5cm,
        bbllx=220pt,bblly=480pt,bburx=375pt,bbury=763pt}
\caption{\it Residual IR poles proportional to $\zeta(2)$ in the interference 
amongst one-loop box diagrams. }
\label{boxbyboxz2}
\ec
\end{figure}

\begin{figure}[h]
\bc
\epsfig{file=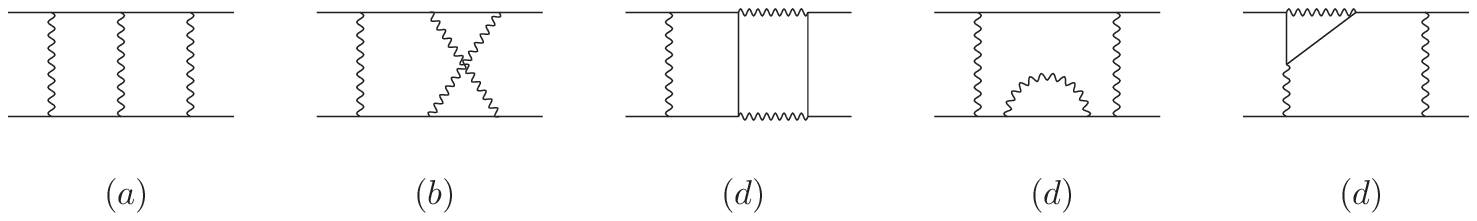,height=2.5cm,width=4.5cm,
        bbllx=220pt,bblly=680pt,bburx=375pt,bbury=763pt}
\caption{\it Two-loop photonic box topologies.}
\label{fig2lbox}
\ec
\end{figure}

In the total differential cross section, the residual poles of Fig.~\ref{figz2s2}
cancel  themselves out. Combining the various contributions shown in 
Fig.~\ref{figz2st}, the residual IR poles do not cancel. Residual single poles
proportional to $\zeta(2)$ also arise in the combinations shown in
Figs.~\ref{figz23334} and \ref{boxbyboxz2}. Clearly, such residual poles  cancel 
once the contribution of the two-loop box graphs is included in 
the cross section at order $\alpha^4$. The five two-loop photonic box topologies
are shown in Fig.~\ref{fig2lbox}.
%

%


\section{Logarithmic Expansions \label{LL}}

In order to check our calculations against the results available in the 
literature, the contributions to the cross section at order $\alpha^4$
described in the previous sections are expanded  in the limit in which 
the mass of the electron is small with respect to all of the Mandelstam 
invariants $s$, $t$, and $u$\footnote{Note that this expansion is not valid
for very small scattering angle, corresponding to $|t| < m^2$, and for 
almost-backward scattering, corresponding to $|u| < m^2$.}. In this 
limit, it is customary to write the two-loop photonic cross section as 
follows (see \cite{Bas,Penin:2005kf}):
\be
\frac{d \sigma_2^{({\tt ph})}}{d \sigma_0} \equiv 
\sum_{i} \frac{d \sigma_2^{({V,i})} + d \sigma_2^{({S,i})}}{d \sigma_0} = 
\delta_2^{(2)} \ln^2\left(\frac{s}{m^2}\right) + 
\delta_2^{(1)} \ln \left(\frac{s}{m^2}\right) + \delta_2^{(0)} + 
{\mathcal O}\left(\frac{m^2}{s} \right) \, . \label{expph}
\ee
In the equation above, $i ={\tt ph\,\,\,Vertices},{\tt ph\,\,\,Boxes}$. Following the
notation adopted in the previous sections,
\bea
\frac{d \sigma_2^{(V, {\tt ph \,\,\,Vertices})}}{d \Omega} &=& 
\left. \frac{d \sigma_2^{(V, {\tt ph \,\,\,Vertices})}}{d \Omega}\right|_{({\tt ph
\,Irr.\, Ver.})}  + 
\left. \frac{d \sigma_2^{(V, {\tt ph \,\,\,Vertices})}}{d \Omega}\right|_{({\tt ph
\, Red. \, Ver. })}\nn \\ & & +
\left. \frac{d \sigma_2^{(V, {\tt ph \,\,\,Vertices})}}{d \Omega}\right|_{({\tt ph
\, Ver. \, Ver. })}  +
\left. \frac{d \sigma_2^{(V, {\tt ph \,\,\,Vertices})}}{d \Omega}\right|_{({\tt ph
\, Ver. \, Box })}  \, ,
\eea
while $\sigma_2^{(V, {\tt ph \,\,\,Boxes})}$ was introduced in Section~\ref{sec4}. 
The corresponding $\sigma_2^{(S,i)}$ cross sections were obtained by pairing 
virtual and soft-photon emission contributions as described
in Section~\ref{IRcancel}. In several points of our discussion, we stressed the 
fact that a non-approximated calculation of the contribution of the two-loop 
photonic box diagrams to the cross section is still missing. However, all of the 
coefficients $\delta_2^{(i)}$ ($i=2,1,0$) in the expansion in Eq.~(\ref{expph}) 
are completely known; the first two can be found in \cite{Bas}, while $\delta_2^{(0)}$ was 
recently obtained in \cite{Penin:2005kf}.  Therefore, by employing the $m^2/s \to 0$ 
limit of the results presented here in combination with \cite{Bas,Penin:2005kf}, 
it is possible to indirectly obtain the $m^2/s \to 0$ limit of the contribution 
of the yet unknown two-loop photonic boxes (and corresponding soft-photon 
emission corrections). In Appendix~\ref{AppB}, we report the expression of such 
a contribution, both before and after adding the corresponding soft-photon
corrections. Moreover, the expansions in the $m^2/s \to 0$ limit of all 
the contributions to the Bhabha scattering cross section discussed in the 
present paper can be found in \cite{file}.

It is known that the small-angle Bhabha scattering cross section is completely 
determined by the Dirac vertex form factor \cite{Fadin:1993ha}. In particular, 
one finds that for the virtual cross section
\be
\frac{d \sigma_2^{(V,{\tt ph})}}{d \sigma_0} \stackrel{\theta \to 0}{=} 
\,\, 6 \left(F_1^{(1l)}(t)\right)^2 + 4 
F_1^{(2l,{ ph})}(t)   \, , 
\ee
where $F_1^{(1l)}$  and $F_1^{(2l,ph)}$ are the UV-renormalized vertex form 
factors already employed in this paper. The IR poles present in the form factors 
are easily removed by adding the soft emission contributions. By introducing 
the IR-finite form factors
\bea
\tilde{F}_1^{(1l)}(t) &=& F_1^{(1l)}(t) + J_{11} + J_{13} \, , \nonumber \\
\tilde{F}_1^{(2l,{ ph})}(t) &=& F_1^{(2l,{ ph})}(t) + \frac{1}{2}
\left(J_{11} + J_{13}\right)^2 \, , 
\eea
one finds that
\bea
\frac{d \sigma_2^{({\tt ph})}}{d \sigma_0} &\stackrel{\theta \to 0}{=}& 
6 \left(\tilde{F}_1^{(1l)}(t)\right)^2 + 4 
\tilde{F}_1^{(2l,{ph})}(t) \, , \nonumber \\
&=& \!\!\! \frac{1}{(1-\xi+\xi^2)^2} \Biggl\{ \ln^2\left(\frac{s}{m^2}\right) \left[ \frac{9}{2}+
           2 \ln^2\left(\frac{4 \omega^2}{s}\right)   
          + 6 \ln\left(\frac{4 \omega^2}{s}\right) \right] 
	  \nn \\
	  &+&\!\!\! \ln\left(\frac{s}{m^2}\right) \Biggl[6 \zeta\left(3\right) -
	  3 \zeta\left(2\right) - \frac{93}{8}
           + 9 \ln\left(\xi\right) 
          - 4 \ln^2\left(\frac{4 \omega^2}{s}\right) \left[1- \ln\left(\xi\right)\right]
          \nn \\
	  &-&\!\!\! 2 \ln\left(\frac{4 \omega^2}{s}\right) \left[7-6 \ln\left(\xi\right) \right]
\Biggr] - 
           9 \zeta\left(3\right)
          + \frac{51}{4} \zeta\left(2\right) 
          - 12 \zeta\left(2\right) \ln\left(2\right)
          - \frac{32}{5} \zeta^2\left(2\right)\nn \\
	  &+&\!\!\!\frac{27}{2} 
          + 6 \zeta\left(3\right) \ln\left(\xi\right) 
          - 3 \zeta\left(2\right) \ln\left(\xi\right)
          - \frac{93}{8} \ln\left(\xi\right)
          + \frac{9}{2} \ln\left(\xi\right)^2 \nn \\
          &+&\!\!\!   \ln^2\left(\frac{4 \omega^2}{s}\right) \left[2 -4 \ln\left(\xi\right)
          + 2 \ln^2\left(\xi\right) \right] \nonumber \\
	  &+&\!\!\!\ln\left(\frac{4 \omega^2}{s}\right) \left[ 8
          - 14 \ln\left(\xi\right) 
          + 6 \ln^2\left(\xi\right) \right] + {\mathcal O}\left(\xi\right)
	  \Biggr\} \, .
\label{smallangle}
\eea
The variable $\xi$ is defined as
\be
\xi = \frac{1 -\cos{\theta}}{2} \, ,
\ee
with $\theta$ the scattering angle in the c. m. frame.
By expanding the exact photonic corrections involving vertex diagrams in the 
$\theta \to 0$ limit and by neglecting terms proportional to the electron mass,
\begin{figure}
\bc
\begin{picture}(0,0)%
\includegraphics{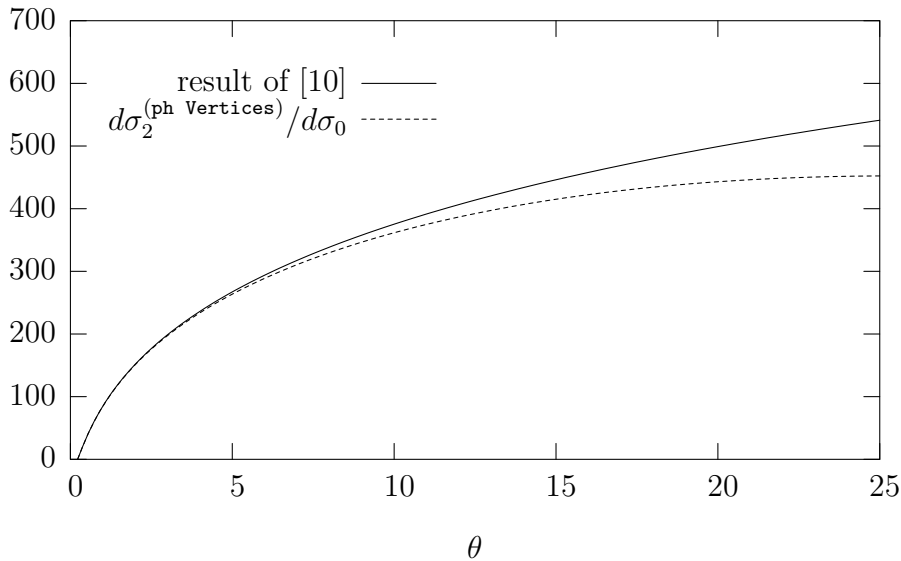}%
\end{picture}%
\setlength{\unitlength}{0.0200bp}%
\begin{picture}(18000,10800)(0,0)%
\put(1650,1979){\makebox(0,0)[r]{\strut{} 0}}%
\put(1650,3161){\makebox(0,0)[r]{\strut{} 100}}%
\put(1650,4342){\makebox(0,0)[r]{\strut{} 200}}%
\put(1650,5524){\makebox(0,0)[r]{\strut{} 300}}%
\put(1650,6705){\makebox(0,0)[r]{\strut{} 400}}%
\put(1650,7887){\makebox(0,0)[r]{\strut{} 500}}%
\put(1650,9068){\makebox(0,0)[r]{\strut{} 600}}%
\put(1650,10250){\makebox(0,0)[r]{\strut{} 700}}%
\put(1925,1429){\makebox(0,0){\strut{} 0}}%
\put(4975,1429){\makebox(0,0){\strut{} 5}}%
\put(8025,1429){\makebox(0,0){\strut{} 10}}%
\put(11075,1429){\makebox(0,0){\strut{} 15}}%
\put(14125,1429){\makebox(0,0){\strut{} 20}}%
\put(17175,1429){\makebox(0,0){\strut{} 25}}%
\put(9550,275){\makebox(0,0){\strut{} $\theta$ }}%
\put(7140,9068){\makebox(0,0)[r]{\strut{}result of \cite{Penin:2005kf}}}%
\put(7140,8393){\makebox(0,0)[r]{\strut{}$d \sigma_{2}^{({\tt ph \,\,\, Vertices})}/ d \sigma_{0}$}}%
\end{picture}%
\caption{\it{Order $\alpha^4$ Bhabha scattering differential cross section 
divided by the cross section in Born approximation, as a function of the 
scattering angle $\theta$. The continuous line represents the result of 
\cite{Penin:2005kf}, while the dashed line represents the corrections involving
at least a vertex graph. The beam energy is chosen equal to 
$0.5$ {\rm GeV} and the soft-photon energy cut-off $\omega$ is set equal to
$E$.}}
\label{figsmalang}
\ec
\end{figure}
we recover  the expression in Eq.~(\ref{smallangle}), which agrees
with \cite{Penin:2005kf} and the theorem in \cite{Fadin:1993ha}.
Consequently, this represents a non-trivial test of our calculation.
As expected, we observe that the interference between one-loop box diagrams 
and one-loop vertex corrections does not contribute to the small-angle cross 
section. The interference of one-loop 
box diagrams amongst themselves has a non-zero small angle limit, and all of
the residual terms are proportional to $\zeta(2)$. These residual 
terms cancel out once the contribution of the two-loop photonic boxes is added;
for this reason they are excluded from the present discussion. The agreement with the results of \cite{Penin:2005kf} 
is clarified in Fig.~\ref{figsmalang}, where we plot
as a function of the scattering angle 
$\theta$ the result of 
\cite{Penin:2005kf} and the corrections of order $\alpha^4$ 
originating from vertex
graphs (Sections \ref{sec1}--\ref{secN}, plus corresponding soft emission
contributions). 
It is easily seen that, at small angles, the vertex corrections completely
determine the cross section. 

In addition, the expansion of the interference of one-loop boxes provides another strong test
of both our calculation and the one discussed in \cite{Penin:2005kf}. 
The interference of some pairs of the diagrams in Fig.~\ref{fig5} gives origin,
in the ratio $d \sigma_2 / d \sigma_0$, to terms proportional to
\be
\frac{\ln^n{\xi}}{(1-\xi)^m} \, , \qquad n = 1,\cdots, 4 \, ; \quad m = 1,2 \, .
\ee
It is possible to observe that, in the sum of all the one-loop box
 interferences,
such terms cancel out, and that 
they do not appear in the complete photonic cross section
at order $\alpha^4$ \cite{Penin:2005kf}. 

\begin{figure}
\bc
\begin{picture}(0,0)%
\includegraphics{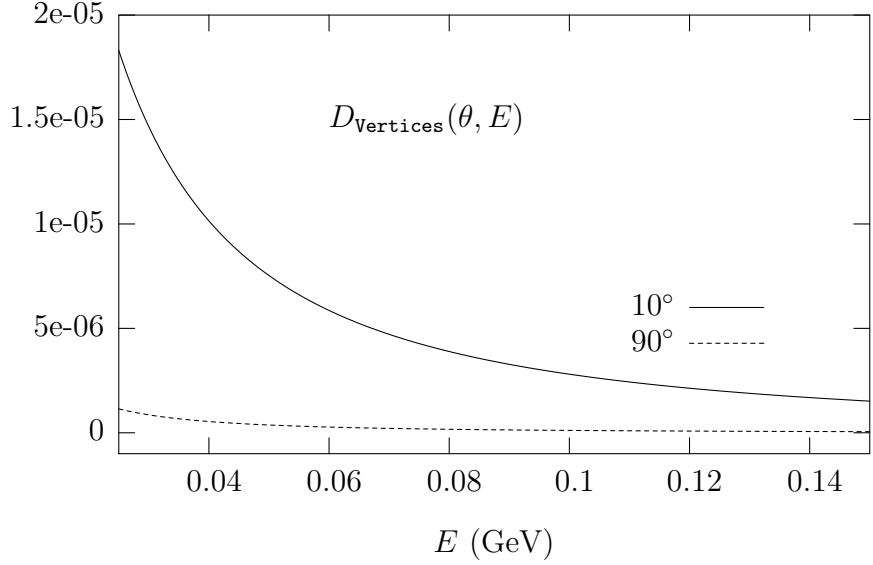}%
\end{picture}%
\setlength{\unitlength}{0.0200bp}%
\begin{picture}(18000,10800)(0,0)%
\put(2750,2373){\makebox(0,0)[r]{\strut{} 0}}%
\put(2750,4342){\makebox(0,0)[r]{\strut{} 5e-06}}%
\put(2750,6311){\makebox(0,0)[r]{\strut{} 1e-05}}%
\put(2750,8281){\makebox(0,0)[r]{\strut{} 1.5e-05}}%
\put(2750,10250){\makebox(0,0)[r]{\strut{} 2e-05}}%
\put(4723,1429){\makebox(0,0){\strut{} 0.04}}%
\put(6987,1429){\makebox(0,0){\strut{} 0.06}}%
\put(9251,1429){\makebox(0,0){\strut{} 0.08}}%
\put(11515,1429){\makebox(0,0){\strut{} 0.1}}%
\put(13779,1429){\makebox(0,0){\strut{} 0.12}}%
\put(16043,1429){\makebox(0,0){\strut{} 0.14}}%
\put(10100,275){\makebox(0,0){\strut{}$E$ (GeV)}}%
\put(6987,8281){\makebox(0,0)[l]{\strut{}$D_{\tt Vertices}(\theta, E) $}}%
\put(13504,4736){\makebox(0,0)[r]{\strut{}$10^{\circ}$}}%
\put(13504,4061){\makebox(0,0)[r]{\strut{}$90^{\circ}$}}%
\end{picture}%
\caption{\it{$D_{\tt Vertices}$ as a function of the beam energy, for 
$\theta = 10^{\circ}$ (solid line) and $\theta = 90^{\circ}$ (dashed line). 
The soft-photon energy cut-off is set equal to $E$}.}
\label{deltaV}
\ec
\end{figure}

\begin{figure}
\bc
\begin{picture}(0,0)%
\includegraphics{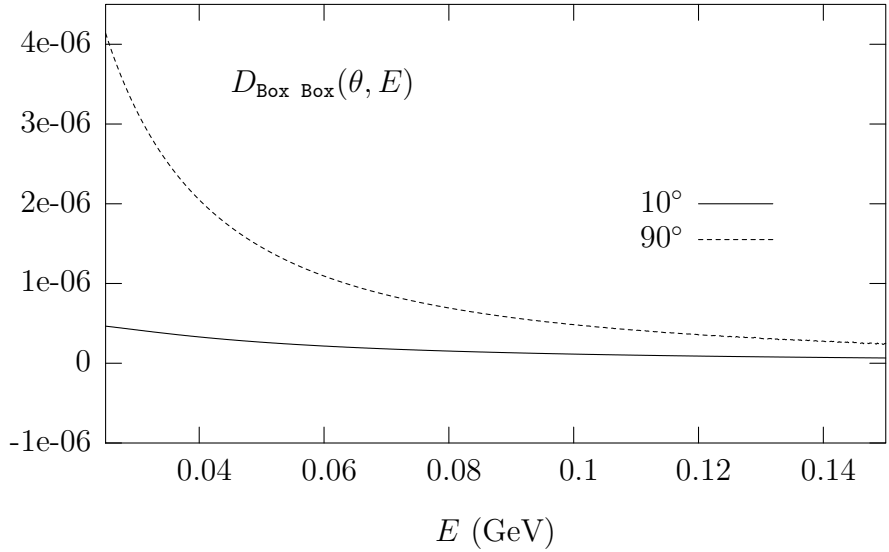}%
\end{picture}%
\setlength{\unitlength}{0.0200bp}%
\begin{picture}(18000,10800)(0,0)%
\put(2200,1979){\makebox(0,0)[r]{\strut{}-1e-06}}%
\put(2200,3483){\makebox(0,0)[r]{\strut{} 0}}%
\put(2200,4987){\makebox(0,0)[r]{\strut{} 1e-06}}%
\put(2200,6490){\makebox(0,0)[r]{\strut{} 2e-06}}%
\put(2200,7994){\makebox(0,0)[r]{\strut{} 3e-06}}%
\put(2200,9498){\makebox(0,0)[r]{\strut{} 4e-06}}%
\put(4239,1429){\makebox(0,0){\strut{} 0.04}}%
\put(6591,1429){\makebox(0,0){\strut{} 0.06}}%
\put(8943,1429){\makebox(0,0){\strut{} 0.08}}%
\put(11295,1429){\makebox(0,0){\strut{} 0.1}}%
\put(13647,1429){\makebox(0,0){\strut{} 0.12}}%
\put(15999,1429){\makebox(0,0){\strut{} 0.14}}%
\put(9825,275){\makebox(0,0){\strut{}$E$ (GeV)}}%
\put(4827,8746){\makebox(0,0)[l]{\strut{}$D_{\tt Box \,\,\, Box}(\theta, E)$}}%
\put(13372,6490){\makebox(0,0)[r]{\strut{}$10^{\circ}$}}%
\put(13372,5815){\makebox(0,0)[r]{\strut{}$90^{\circ}$}}%
\end{picture}%
\caption{\it{$D_{\tt Box \,\,\, Box}$ as a function of the energy, 
for $\theta = 10^{\circ}$ (solid line) and $\theta = 90^{\circ}$ 
(dashed line). The soft-photon energy cut-off is set equal to 
$E$}.}
\label{deltaBB} 
\ec
\end{figure}

The photonic corrections to the Bhabha scattering cross section are now 
known up to terms of ${\mathcal O}(m^2/s)$ excluded; it is possible to use the 
corrections calculated exactly in this work to 
estimate the relevance of the ${\mathcal O}(m^2/s)$ terms.
We define
\be
\frac{d \sigma_2^{(i,{\tt ph})}}{d \Omega} = 
\frac{d \sigma_2^{({V,i})} + d \sigma_2^{({S,i})}}{d \Omega} =
\left. \frac{d \sigma_2^{(i,{\tt ph})}}{d \Omega} \right|_L + 
{\mathcal O}\left(\frac{m^2}{s},\frac{m^2}{t},\frac{m^2}{u}   \right) \, ,
\label{LLexpansion}
\ee 
where the index $i={\tt Vertices}, {\tt Box \, Box}$ represents the contributions 
discussed in Sections~\ref{sec1}--\ref{secN}, and \ref{sec4}, respectively. 
In Figs.~\ref{deltaV} and \ref{deltaBB}, we plot, as a function of the beam energy, 
the quantities
\be
D_{i} = \left(\frac{\alpha}{\pi}\right)^2 \left|\left( 
\frac{d \sigma_2^{(i,{\tt ph})}}{d \Omega} -
\left. \frac{d \sigma_2^{(i,{\tt ph})}}{d \Omega} \right|_L 
\right) \right| \left(\frac{d \sigma_0}{d \Omega} +
\left(\frac{\alpha}{\pi}\right) \frac{d \sigma_1}{d \Omega}
\right)^{-1}
\, .
\ee
%
%
%
%
%
It can be seen from these plots that the terms proportional to the electron mass become
negligible for values of the beam energy that are very small with 
respect to the ones encountered in practically all of the $e^+  e^-$ experiments. 
It is also reasonable to expect that the terms 
proportional to the electron mass are negligible
in the corrections due to the two-loop photonic boxes. In this sense, the approximated 
cross section obtained in \cite{Bas,Penin:2005kf,us2,us3} should be sufficient for all 
phenomenological studies.

\begin{figure}
\bc
\begin{picture}(0,0)%
\includegraphics{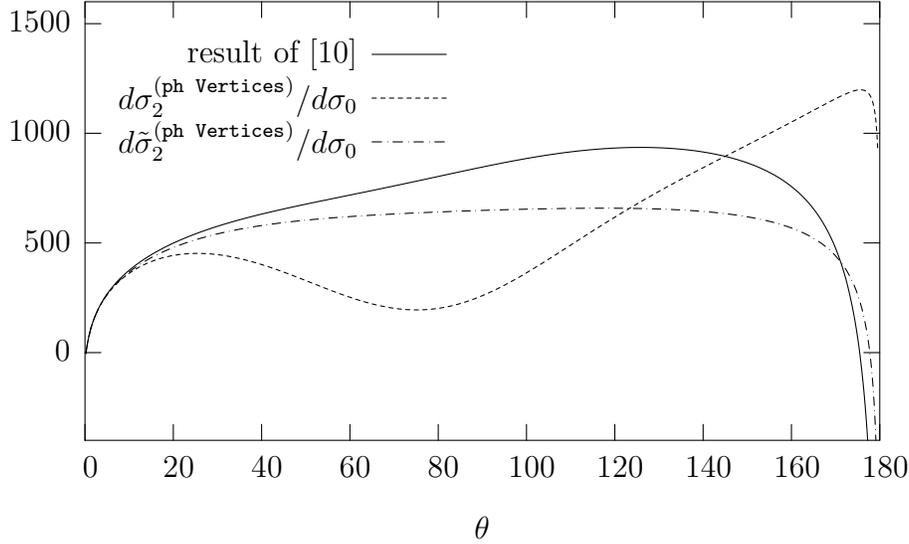}%
\end{picture}%
\setlength{\unitlength}{0.0200bp}%
\begin{picture}(18000,10800)(0,0)%
\put(1925,3633){\makebox(0,0)[r]{\strut{} 0}}%
\put(1925,5701){\makebox(0,0)[r]{\strut{} 500}}%
\put(1925,7769){\makebox(0,0)[r]{\strut{} 1000}}%
\put(1925,9836){\makebox(0,0)[r]{\strut{} 1500}}%
\put(2200,1429){\makebox(0,0){\strut{} 0}}%
\put(3864,1429){\makebox(0,0){\strut{} 20}}%
\put(5528,1429){\makebox(0,0){\strut{} 40}}%
\put(7192,1429){\makebox(0,0){\strut{} 60}}%
\put(8856,1429){\makebox(0,0){\strut{} 80}}%
\put(10519,1429){\makebox(0,0){\strut{} 100}}%
\put(12183,1429){\makebox(0,0){\strut{} 120}}%
\put(13847,1429){\makebox(0,0){\strut{} 140}}%
\put(15511,1429){\makebox(0,0){\strut{} 160}}%
\put(17175,1429){\makebox(0,0){\strut{} 180}}%
\put(9687,275){\makebox(0,0){\strut{}$\theta$}}%
\put(7333,9257){\makebox(0,0)[r]{\strut{}result of \cite{Penin:2005kf}}}%
\put(7333,8432){\makebox(0,0)[r]{\strut{}$d\sigma_{2}^{({\tt ph \,\,\, Vertices})}/d\sigma_{0}$}}%
\put(7333,7607){\makebox(0,0)[r]{\strut{}$d\tilde{\sigma}_{2}^{({\tt ph \,\,\, Vertices})}/d\sigma_{0}$}}%
\end{picture}%
\caption{\it{Comparison of the vertex corrections with the complete photonic 
cross section at order $\alpha^4$. $E$ and $\omega$ as in Fig.~\ref{figsmalang}
}.}
\label{vePe}
\ec
\end{figure}

\begin{figure}
\bc
\begin{picture}(0,0)%
\includegraphics{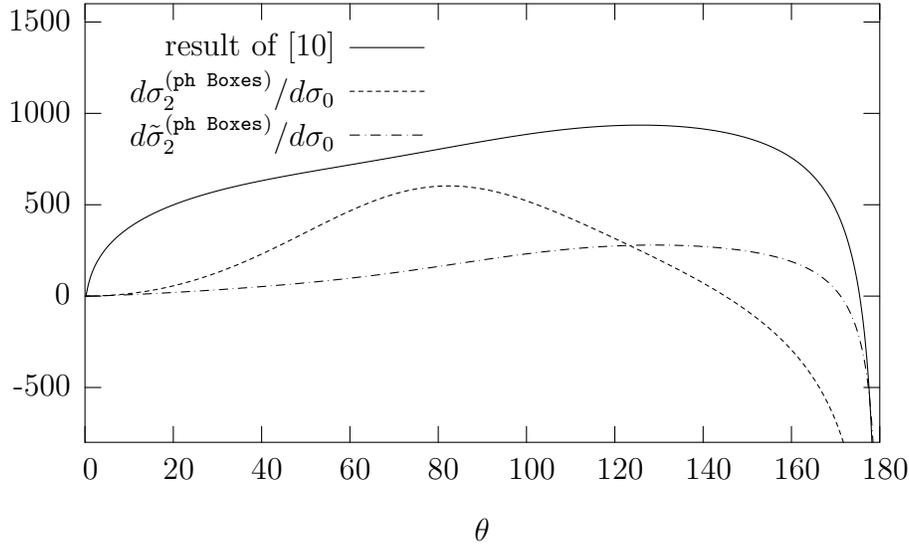}%
\end{picture}%
\setlength{\unitlength}{0.0200bp}%
\begin{picture}(18000,10800)(0,0)%
\put(1925,3013){\makebox(0,0)[r]{\strut{}-500}}%
\put(1925,4736){\makebox(0,0)[r]{\strut{} 0}}%
\put(1925,6459){\makebox(0,0)[r]{\strut{} 500}}%
\put(1925,8182){\makebox(0,0)[r]{\strut{} 1000}}%
\put(1925,9905){\makebox(0,0)[r]{\strut{} 1500}}%
\put(2200,1429){\makebox(0,0){\strut{} 0}}%
\put(3864,1429){\makebox(0,0){\strut{} 20}}%
\put(5528,1429){\makebox(0,0){\strut{} 40}}%
\put(7192,1429){\makebox(0,0){\strut{} 60}}%
\put(8856,1429){\makebox(0,0){\strut{} 80}}%
\put(10519,1429){\makebox(0,0){\strut{} 100}}%
\put(12183,1429){\makebox(0,0){\strut{} 120}}%
\put(13847,1429){\makebox(0,0){\strut{} 140}}%
\put(15511,1429){\makebox(0,0){\strut{} 160}}%
\put(17175,1429){\makebox(0,0){\strut{} 180}}%
\put(9687,275){\makebox(0,0){\strut{}$\theta$}}%
\put(6917,9423){\makebox(0,0)[r]{\strut{}result of \cite{Penin:2005kf}}}%
\put(6917,8598){\makebox(0,0)[r]{\strut{}$d\sigma_{2}^{({\tt ph \,\,\, Boxes})}/d\sigma_{0}$}}%
\put(6917,7773){\makebox(0,0)[r]{\strut{}$d\tilde{\sigma}_{2}^{({\tt ph \,\,\, Boxes})}/d\sigma_{0}$}}%
\end{picture}%
\caption{\it{Comparison of the box corrections with the complete photonic 
cross section at order $\alpha^4$. $E$ and $\omega$ as in Fig.~\ref{figsmalang}.}}
\label{boPe}
\ec
\end{figure}

Finally, in Figs.~\ref{vePe} and \ref{boPe}, it is possible to compare the various 
contributions to the Bhabha scattering differential cross section known at
present with the complete photonic cross section in the $m^2/s \to 0$ limit
\cite{Penin:2005kf}. The dashed line in Fig.~\ref{vePe} corresponds to the 
contribution to the cross section defined in Eq.~(\ref{LLexpansion}) for 
$i={\tt Vertices}$, plotted as a function of the scattering angle. The local minimum
at $\theta \sim 80^{\circ}$ and the maximum in the backward direction are due to
spurious terms proportional to the monomials
\be
\zeta(2) \ln^{2} \left( \frac{s}{m^2} \right) \, , \quad
\zeta(2) \ln \left( \frac{s}{m^2} \right) \ln \xi \, , \quad \mbox{and}
\quad \zeta(2) \ln \left( \frac{s}{m^2} \right) \ln (1-\xi) \, .
\label{spurious}
\ee
These terms are not present in the complete cross section (they cancel out against 
analogous contributions deriving from the interferences of two-loop box diagrams
with the tree-level amplitude and from the interference  one-loop box diagrams amongst themselves). 
Removing them from the vertex corrections, one obtains the dashed-dotted curve in 
Fig.~\ref{vePe}, which is smoother than the dashed one. The solid curve represents 
the complete photonic cross section. As is already observed above, the vertex 
contribution reproduces  the full result in the small-angle region.

The dashed line in Fig.~\ref{boPe} represents the quantity
\be
\frac{d \sigma_2^{({\tt ph\,\,\, Boxes})}}{d \sigma_0}  = 
\frac{d \sigma_2^{({\tt ph})}}{d \sigma_0} - 
\frac{d \sigma_2^{({\tt ph\,\,\,Vertices})}}{d \sigma_0} = 
\frac{d \sigma_2^{({\tt ph\, Box \,Box})}}{d \sigma_0} +
\frac{d \sigma_2^{({\tt ph\, 2L \, Box})}}{d \sigma_0} \, .
\ee
%
%
In the equation above, $\sigma_2^{({\tt ph\,  Box \, Box})}$ was introduced 
in Eq.~(\ref{LLexpansion}), while $\sigma_2^{({\tt ph\, 2L \, Box})}$ is 
the contribution to the cross section of the two-loop photonic boxes interfered 
with the tree-level amplitude. The expression of the latter 
is given in Appendix~\ref{AppB}.
Also in this case, the behavior at $\theta \sim 80^{\circ}$ and at 
$\theta \sim 180^{\circ}$ is dominated by the spurious terms 
of the kind of Eq.~(\ref{spurious});
by removing them, we obtain the dashed-dotted curve.

\begin{figure}
\bc
\begin{picture}(0,0)%
\includegraphics{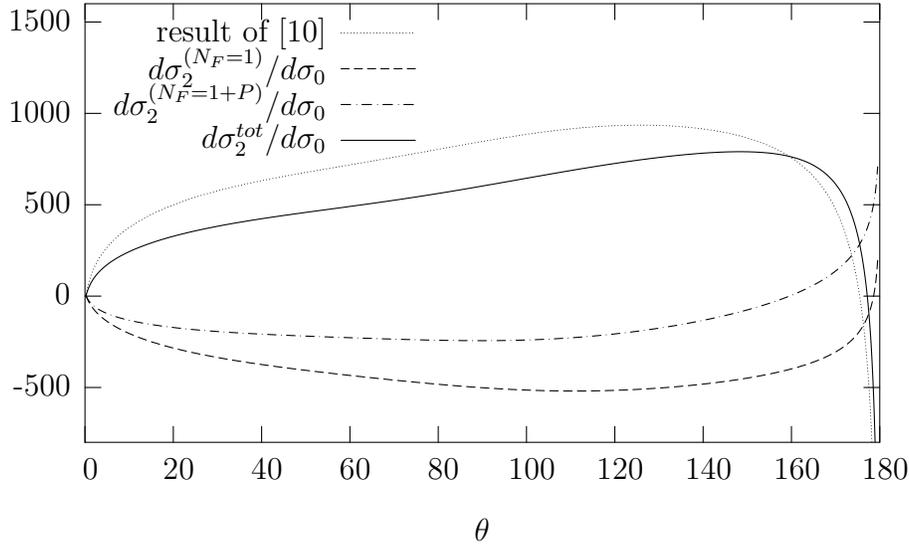}%
\end{picture}%
\setlength{\unitlength}{0.0200bp}%
\begin{picture}(18000,10800)(0,0)%
\put(1925,3013){\makebox(0,0)[r]{\strut{}-500}}%
\put(1925,4736){\makebox(0,0)[r]{\strut{} 0}}%
\put(1925,6459){\makebox(0,0)[r]{\strut{} 500}}%
\put(1925,8182){\makebox(0,0)[r]{\strut{} 1000}}%
\put(1925,9905){\makebox(0,0)[r]{\strut{} 1500}}%
\put(2200,1429){\makebox(0,0){\strut{} 0}}%
\put(3864,1429){\makebox(0,0){\strut{} 20}}%
\put(5528,1429){\makebox(0,0){\strut{} 40}}%
\put(7192,1429){\makebox(0,0){\strut{} 60}}%
\put(8856,1429){\makebox(0,0){\strut{} 80}}%
\put(10519,1429){\makebox(0,0){\strut{} 100}}%
\put(12183,1429){\makebox(0,0){\strut{} 120}}%
\put(13847,1429){\makebox(0,0){\strut{} 140}}%
\put(15511,1429){\makebox(0,0){\strut{} 160}}%
\put(17175,1429){\makebox(0,0){\strut{} 180}}%
\put(9687,275){\makebox(0,0){\strut{} $\theta$ }}%
\put(6750,9699){\makebox(0,0)[r]{\strut{}result of \cite{Penin:2005kf}}}%
\put(6750,9024){\makebox(0,0)[r]{\strut{}$d\sigma_{2}^{(N_{F}=1)}/ d\sigma_{0}$}}%
\put(6750,8349){\makebox(0,0)[r]{\strut{}$d\sigma_{2}^{(N_{F}=1+P)}/ d\sigma_{0}$}}%
\put(6750,7674){\makebox(0,0)[r]{\strut{}$d\sigma_{2}^{tot}/ d\sigma_{0}$}}%
\end{picture}%
\caption{\label{all} \it{Photonic, $N_F=1$, and total contributions to the 
cross section at order $\alpha^4$. $E$ and $\omega$ as in Fig.~\ref{figsmalang}.
The pair production cut-off is set equal to $\omega$.}}
\ec
\end{figure}

For completeness, in Fig.~\ref{all} we plot the photonic \cite{Penin:2005kf}
and $N_F=1$ \cite{us3} contributions to the Bhabha scattering cross section 
at order $\alpha^4$. The dotted line represents the photonic corrections.
The corrections of ${\mathcal O}(\alpha^4 (N_F=1))$ (dashed line) have, for 
this choice of $\omega$ ($\omega = E$), an opposite sign with respect to the 
photonic corrections. However, it is necessary to say that the  ${\mathcal O}(\alpha^4 (N_F
=1))$ of \cite{us3} include large contributions proportional to $\ln^3(s/m^2)$ 
that cancel out once the contribution of the soft pair production is included. 
Of the latter, only the terms proportional to $\ln^n(s/m^2)$ ($n=1,2,3$) are 
known (see \cite{Arbuzov:1995vj}); we checked  that
they cancel the $\ln^3(s/m^2)$ term of the ${\mathcal O}(\alpha^4 (N_F
=1))$ cross section. The dashed-dotted line represents the sum of the
${\mathcal O}(\alpha^4 (N_F=1))$ cross section with the known terms of the pair production
corrections\footnote{The pair production, calculated in \cite{Arbuzov:1995vj}
up to terms enhanced by $\ln^{n}(s/m^2)$ included, depends upon 
a cut-off on the energy of the soft electron-positron pair. In the numerical 
evaluation of Fig.~\ref{all} we set
\be
\ln({\mathrm{D}}) = \frac{1}{2} \ln \left( \frac{4 \Omega^2}{s} \right) \, ,
\ee
with $\ln({\mathrm{D}})$ defined in \cite{Arbuzov:1995vj} and $\Omega$  
numerically equal to the soft-photon cut-off: $\Omega = \omega$.}. The solid line is the 
complete order $\alpha^4$ QED Bhabha scattering cross section, including photonic, 
$N_F=1$ and pair production contributions.

\section{Conclusions\label{summ}}

In the present paper, we obtained analytic non approximated expressions for all
of  the 
photonic corrections to the QED Bhabha scattering differential cross section at 
second order ($\mathcal{O}(\alpha^4)$), except for the ones 
deriving from the interference 
of two-loop photonic box diagrams with the tree-level amplitude; 
at present, the integrals 
necessary to the calculation of the latter are not known. The calculations were 
carried out by retaining the full dependence on the electron mass $m$. The results 
are valid for arbitrary values of the c.~m. energy $s$ and momentum transfer $t$.

We included a discussion of the soft-photon emission at order $\alpha^4$,
employing dimensional regularization to handle the IR-divergent terms. We proved
that the term proportional to $(D-4)$ in the Laurent expansion of the photon
phase-space integral $I_{1j}$ in Eq.~(\ref{Iij}) does not contribute to the Bhabha
scattering cross section.

After subtracting the IR singularities by adding the contribution
of the soft-photon emission graphs,  we expanded the contributions to
the cross section discussed in Sections~\ref{sec1}--\ref{sec4} in the $m^2/s
\to 0$ limit.  In this way, it was possible to cross-check large parts of the
result of \cite{Penin:2005kf}, as well as to efficiently test our calculations.  
By subtracting the contributions to the cross section obtained in this paper from
the cross section of \cite{Penin:2005kf}, it was also possible to indirectly
obtain the contribution due to the interference of the two-loop photonic
diagrams with the tree-level amplitude, up to terms suppressed by positive
powers of the electron mass. 

By comparing the non approximated results with the corresponding $m^2/s \to 0$
limit, we explicitly checked that the contribution  to the cross section of 
the terms proportional to positive powers of the ratio
$m^2/s$ is negligible at high- and intermediate-energy $e^+ e^-$ colliders. It
is reasonable that the same conclusion applies to the contribution to the order
$\alpha^4$ cross section involving the two-loop photonic box diagrams.
For what concerns phenomenological studies, the results of 
\cite{Bas,Penin:2005kf,us2,us3} therefore provide a complete
expression of the virtual and soft-photon emission corrections to the Bhabha 
scattering cross section at order $\alpha^4$. 
Nevertheless, a future calculation of the two-loop 
photonic box graphs for $m \neq 0$ would represent a very interesting result 
in the field of multi-loop calculations.  \\

\vspace*{5mm}

\noindent {\large{\bf Acknowledgments}} \\

\noindent The authors wish to thank H. Czyz, B, Tausk, and L. Trentadue for 
useful discussions and suggestions, as well as A. Penin for cross-checks, 
discussions, and for providing us with the expression of Eq.~(3) of 
\cite{Penin:2005kf} in electronic format. 
We are grateful to J.~Vermaseren for his kind assistance in the use of the 
algebra manipulating program {\tt FORM}~\cite{FORM}.

\appendix

\boldmath
\section{$C_{i j}$ Functions \label{AppA}} 
\unboldmath

In this appendix, we collect the explicit expressions of the $c_{ij}$ functions
employed in the main text:
\bea
c_{11}(s,t) &=&  (s+t)^2 -4 m^2 \, , \\
c_{12}(s,t) &=& \frac{1}{2} \left[(s+t)^2 - s t -2 m^2 (s+t)\right] \, , \\
c_{13}(s,t) &=& 2 \left( s t - \frac{3}{2} t m^2 + \frac{3}{4} t^2\right)\, ,\\
c_{14}(s,t) &=& - \frac{1}{4} s t \, , \\
c_{15}(s,t) &=& s t + \frac{s t^2}{4 m^2} - 2 t m^2 + \frac{3}{4} t^2 + 
\frac{t^3}{8 m^2} \, , \\
c_{16}(s,t) &=& \frac{1}{4} \left(s t  -4 t m^2 + 2 t^2 \right) \, , \\
c_{17}(s,t) &=& \frac{1}{4} s t \left(1 +  \frac{t}{m^2} 
+ \frac{s}{m^2} \right) \, , \\
c_{21}(s,t) &=& 2 \left[ (s-2 m^2)^2 + s t +\frac{t^2}{2} \right] \, , \\
c_{22}(s,t) &=& \frac{t^2}{2} \, , \\
c_{23}(s,t) &=& t \left(t + 2 m^2\right) \, , \\
c_{24}(s,t) &=& 0 \, , \\
c_{25}(s,t) &=& \frac{3 t^2}{2} \, , \\
c_{26}(s,t) &=& \frac{t^2}{2} \, , \\
c_{27}(s,t) &=& s t - \frac{s t^2}{4 m^2} -\frac{s^2 t}{4 m^2}  
+ \frac{3}{4} t^2 \, .
\eea

\boldmath
\section{Photonic Double Boxes \label{AppB}} 
\unboldmath

We provide here the expansion of the interference of the two-loop photonic boxes 
with the tree-level amplitude in the limit $m^2/s \to 0$. Below one can find
the contribution of the two-loop photonic boxes to the virtual cross section
as well as the same contribution after the subtraction of the corresponding
soft-photon corrections. \\

\noindent {\bf Without including soft radiation} \\

\bea
\frac{d \sigma_{2}^{(V,{\tt ph\, 2L \, Box})}}{d \sigma_0} &=& 
              \delta_{(2)}^{(V,{\tt 2L \, Box},3)} \log^{3}{\left( \frac{s}{m^2} \right)}
            + \delta_{(2)}^{(V,{\tt 2L \, Box},2)} \log^{2}{\left( \frac{s}{m^2} \right)}
            + \delta_{(2)}^{(V,{\tt 2L \, Box},1)} \log{\left( \frac{s}{m^2} \right)}
	    \nn\\
& & 
	    + \delta_{(2)}^{({\tt 2L \, Box},0)} + {\mathcal O} \left(
	    \frac{m^2}{s} \right) \, ,
\eea
where
\bea
\delta_{(2)}^{(V,{\tt 2L \, Box},3)} \! \! & = & \! \! \frac{1}{(1-\xi+\xi^2)^2}   \Biggl\{ 
       -    \Bigl(
            \frac{7}{3}
          - \frac{14}{3} \xi
          + 7 \xi^2
          - \frac{14}{3} \xi^3
          + \frac{7}{3} \xi^4
          \Bigr) \ln(1 - \xi)
       -    \Bigl(
            \frac{7}{6} \xi
          - \frac{7}{2} \xi^2 \nn\\
\! \! & & \! \! 
          + \frac{7}{2} \xi^3
          - \frac{7}{3} \xi^4
          \Bigr) \ln(\xi) \Biggr\} \, , \\
\delta_{(2)}^{(V,{\tt 2L \, Box},2)} \! \! & = & \! \! \frac{1}{(1-\xi+\xi^2)^2}   \Bigl\{ \nn\\
\! \! & & \! \! 
 \frac{1}{(D-4)} \, \Bigl[
         ( 
          - 6
          + 12 \xi
          - 18 \xi^2
          + 12 \xi^3
          - 6 \xi^4
          ) \ln(1 - \xi) 
       -    (
            3 \xi
          - 9 \xi^2
          + 9 \xi^3 \nn\\
\! \! & & \! \! \hspace*{15mm}
          - 6 \xi^4
          ) \ln(\xi) \Bigr] \nn\\
\! \! & & \! \! 
       -    \Bigl(
            12
          - \frac{27}{2} \xi \! 
          + \!  \frac{63}{4} \xi^2
          - 6 \xi^3
          \Bigr) \zeta(2) \! 
       +   \! \Bigl(
            2
          - \frac{19}{4} \xi \! 
          +  \! \frac{27}{4} \xi^2
          - \frac{19}{4} \xi^3 \! 
          +  \! 2 \xi^4
          \Bigr) \ln^2(1 - \xi) \nn\\
\! \! & & \! \! 
       +   \!   \Bigl(
            \frac{11}{2}
          - \frac{29}{4} \xi
          +  \! \frac{21}{2} \xi^2
          - \frac{29}{4} \xi^3
          +  \! \frac{11}{2} \xi^4
          \Bigr) \ln(1 - \xi)
       -    \Bigl(
            7
          - 14 \xi
          + \frac{69}{4} \xi^2
          - \frac{19}{2} \xi^3 \nn\\
\! \! & & \! \! 
          + 4 \xi^4
          \Bigr) \ln(\xi) \ln(1 - \xi)
       +    \Bigl(
            \frac{1}{2} \xi
          - \frac{21}{4} \xi^2
          + \frac{27}{4} \xi^3
          - \frac{11}{2} \xi^4
          \Bigr) \ln(\xi)
       -    \Bigl(
            \frac{23}{8} \xi
          - \frac{45}{8} \xi^2 \nn\\
\! \! & & \! \! 
          + \frac{33}{8} \xi^3
          - 2 \xi^4
          \Bigr) \ln^2(\xi)
      \Biggr\}  \, , \\
\delta_{(2)}^{(V,{\tt 2L \, Box},1)} \! \! & = & \! \! \frac{1}{(1-\xi+\xi^2)^2} \Bigl\{  \nn\\
\! \! & & \! \! 
 \frac{1}{(D-4)^2} \, \Bigl[ 
       -     (
            8
          - 16 \xi
          + 24 \xi^2
          - 16 \xi^3
          + 8 \xi^4
          ) \ln(1 - \xi)
       -     (
            4 \xi
          - 12 \xi^2
          + 12 \xi^3 \nn\\
\! \! & & \! \! \hspace*{15mm}
          - 8 \xi^4
          ) \ln(\xi) \Bigr] \nn\\
\! \! & & \! \! 
+ \frac{1}{(D-4)} \, \Biggl[ 
            (
          - 24
          + 30 \xi
          - 33 \xi^2
          + 12 \xi^3
          )  \zeta(2)
       +     (
            10
          - 15 \xi
          + 22 \xi^2
          - 15 \xi^3 \nn\\
\! \! & & \! \! \hspace*{15mm}
          +  \! 10 \xi^4 
          ) \ln(1 - \xi)
       +   \!   (
            4 \! 
          -  \! 9 \xi \! 
          +  \! 13 \xi^2 \! 
          -  \! 9 \xi^3 \! 
          +  \! 4 \xi^4
          ) \ln^2(1 - \xi) 
       +   \!  (
            2 \xi
          - 11 \xi^2 \nn\\
\! \! & & \! \! \hspace*{15mm}
          +  \! 13 \xi^3
          - 10 \xi^4
          ) \ln(\xi)
       -     (
            12
          - 24 \xi \! 
          +  \! 31 \xi^2
          - 18 \xi^3 \! 
          +  \! 8 \xi^4
          ) \ln(\xi) \ln(1 - \xi) \nn\\
\! \! & & \! \! \hspace*{15mm}
       -     \Bigl(
            \frac{9}{2} \xi
          - \frac{19}{2} \xi^2
          + \frac{15}{2} \xi^3
          - 4 \xi^4
          \Bigr) \ln^2(\xi) \Biggr] \nn\\
\! \! & & \! \! 
       +    \Bigl(
            \frac{27}{2} \xi
          - \frac{33}{2} \xi^2
          + \frac{27}{2} \xi^3
          \Bigr) \zeta(2)
       +    \Bigl(
            \frac{5}{2} \xi
          - 4 \xi^2
          + \frac{5}{2} \xi^3
          \Bigr)
       +    \Bigl(
            12
          - 38 \xi
          + \frac{185}{2} \xi^2 \nn\\
\! \! & & \! \! 
          - 89 \xi^3
          + 46 \xi^4
          \Bigr) \zeta(3) \zeta(2) \ln(1 - \xi)
       +    \Bigl(
            \frac{2}{3} \xi
          - \frac{1}{4} \xi^2
          + \frac{1}{6} \xi^3
          + \frac{1}{3} \xi^4
          \Bigr) \ln^3(1 - \xi) \nn\\
\! \! & & \! \! 
       -    \Bigl(
            \frac{3}{2} \xi
          - \frac{5}{2} \xi^2
          + \frac{3}{2} \xi^3
          \Bigr) \ln^2(1 - \xi)
       -    \Bigl(
            2
          - 3 \xi
          + 3 \xi^3
          - 2 \xi^4
          \Bigr) \ln(1 - \xi) {\mathrm{Li}}_{2}(\xi) \nn\\
\! \! & & \! \! 
       -    \Bigl(
            12 \! 
          -  \! \frac{67}{4} \xi \! 
          +  \! \frac{49}{2} \xi^2 \! 
          -  \! \frac{67}{4} \xi^3 \! 
          +  \! 12 \xi^4
          \Bigr) \ln(1 - \xi) \! 
       -   \!   \Bigl(
            \xi \! 
          -  \! \frac{7}{2} \xi^2 \! 
          +  \! 4 \xi^3 \! 
          -  \! 2 \xi^4
          \Bigr) {\mathrm{Li}}_{3}(1 - \xi) \nn\\
\! \! & & \! \! 
       +   \!   \Bigl(
            2
          - 4 \xi \! 
          +  \! \frac{7}{2} \xi^2
          - \xi^3
          \Bigr) {\mathrm{Li}}_{3}\left( - \frac{\xi}{(1-\xi)} \right)
       -    \Bigl(
            2
          - \frac{5}{2} \xi \! 
          +  \! 3 \xi^2
          - \frac{5}{2} \xi^3 \! 
          +  \! 2 \xi^4
          \Bigr) {\mathrm{Li}}_{3}(\xi) \nn\\
\! \! & & \! \! 
       -    \Bigl(
            24
          - \frac{57}{2} \xi
          + \frac{117}{2} \xi^2
          - \frac{123}{2} \xi^3
          + 48 \xi^4
          \Bigr) \zeta(2) \ln(\xi)
       +    \Bigl(
            3
          - \frac{15}{2} \xi
          + 8 \xi^2
          - \frac{9}{2} \xi^3 \nn\\
\! \! & & \! \! 
          + \xi^4
          \Bigr) \ln(\xi) \ln^2(1 - \xi)
       +    \Bigl(
            11
          - 12 \xi
          + 8 \xi^2
          + \frac{1}{2} \xi^3
          \Bigr) \ln(\xi) \ln(1 - \xi)
       +    \Bigl(
            4
          - \frac{13}{2} \xi \nn\\
\! \! & & \! \! 
          + \frac{13}{2} \xi^2
          - \frac{7}{2} \xi^3
          + 2 \xi^4
          \Bigr) \ln(\xi) {\mathrm{Li}}_{2}(\xi)
       -    \Bigl(
            \frac{5}{4} \xi
          - \frac{49}{4} \xi^2
          + \frac{31}{2} \xi^3
          - 12 \xi^4
          \Bigr) \ln(\xi) \nn\\
\! \! & & \! \! 
       -    \Bigl(
            5
          - \frac{49}{4} \xi \! 
          +  \! 12 \xi^2
          - \frac{13}{4} \xi^3
          - \xi^4
          \Bigr) \ln^2(\xi) \ln(1 - \xi)
       +  \!  \Bigl(
            \frac{7}{8} \xi
          - \frac{5}{4} \xi^2 \! 
          +  \! \frac{7}{8} \xi^3
          \Bigr) \ln^2(\xi) \nn\\
\! \! & & \! \! 
       -    \Bigl(
            \frac{19}{6} \xi
          - \frac{25}{6} \xi^2
          + \frac{3}{2} \xi^3
          + \frac{1}{3} \xi^4
          \Bigr) \ln^3(\xi)
	  \Biggr\} \, , \\
\delta_{(2)}^{(V,{\tt 2L \, Box},0)} \! \! & = & \! \! \frac{1}{(1-\xi+\xi^2)^2} \Bigl\{  \nn\\
\! \! & & \! \! 
 \frac{1}{(D-4)^2} \, \Bigl[ 
       -    (
            24
          - 36 \xi
          + 36 \xi^2
          - 12 \xi^3
          ) \zeta(2)
       +    (
            8
          - 16 \xi
          + 24 \xi^2
          - 16 \xi^3\nn\\
\! \! & & \! \! \hspace*{15mm}
          + 8 \xi^4
          ) \ln(1 - \xi) \! 
       +  \!  (
            4
          - 8 \xi \! 
          +  \! 12 \xi^2
          - 8 \xi^3 \! 
          +  \! 4 \xi^4
          ) \ln^2(1 - \xi) \! 
       +   \!   (
            4 \xi
          - 12 \xi^2 \nn\\
\! \! & & \! \! \hspace*{15mm}
          + 12 \xi^3
          - 8 \xi^4
          ) \ln(\xi)
       -    (
            8
          - 16 \xi
          + 24 \xi^2
          - 16 \xi^3
          + 8 \xi^4
          ) \ln(\xi) \ln(1 - \xi) \nn\\
\! \! & & \! \! \hspace*{15mm}
       -    (
            2 \xi
          - 6 \xi^2
          + 6 \xi^3
          - 4 \xi^4
          ) \ln^2(\xi)
	   \Bigr] \nn\\
\! \! & & \! \! 
+ \frac{1}{(D-4)}  \! \Biggl[  \! 
            (
            12 \xi \! 
          -  \! 15 \xi^2 \! 
          +  \! 12 \xi^3 \! 
          ) \zeta(2) \! 
       +  \!   (
            4 \! 
          -  \! 20 \xi \! 
          +  \! 69 \xi^2 \! 
          -  \! 74 \xi^3 \! 
          +  \! 40 \xi^4
          )  \zeta(2)  \! \ln(1  \! -  \! \xi) \nn\\
\! \! & & \! \! \hspace*{15mm}
       -    (
            8 \! 
          -  \! 11 \xi \! 
          +  \! 16 \xi^2 \! 
          - \!  11 \xi^3 \! 
          +  \! 8 \xi^4
          ) \ln(1 - \xi) \! 
       -    \!  (
            2 \xi \! 
          -  \! 3 \xi^2 \! 
          +  \! 2 \xi^3
          ) \ln^2(1 - \xi) \nn\\
\! \! & & \! \! \hspace*{15mm}
       +   \!   (
            \xi
          - \xi^2 \! 
          +  \! \xi^3
          ) \ln^3(1 - \xi)
       -  (
            24
          - 32 \xi \! 
          +  \! 57 \xi^2
          - 54 \xi^3 \! 
          +  \! 40 \xi^4
          ) \zeta(2) \ln(\xi) \nn\\
\! \! & & \! \! \hspace*{15mm}
       -  (
            \xi
          - 8 \xi^2
          + 10 \xi^3
          - 8 \xi^4
          ) \ln(\xi)
       +   (
            10
          - 11 \xi
          + 8 \xi^2
          ) \ln(\xi) \ln(1 - \xi)  \nn\\
\! \! & & \! \! \hspace*{15mm}
       +   (
            4
          - 9 \xi
          + 8 \xi^2
          - 3 \xi^3
          ) \ln(\xi) \ln^2(1 - \xi) 
       +    \Bigl(
            \frac{1}{2} \xi
          - \frac{3}{2} \xi^2
          + \frac{1}{2} \xi^3
          \Bigr) \ln^2(\xi) \nn\\
\! \! & & \! \! \hspace*{15mm}
       -    \Bigl(
            6
          - \frac{27}{2} \xi
          + \frac{27}{2} \xi^2
          - \frac{9}{2} \xi^3
          \Bigr) \ln^2(\xi) \ln(1 - \xi)
       -    \Bigl(
            \frac{5}{2} \xi
          - \frac{7}{2} \xi^2 \nn\\
\! \! & & \! \! \hspace*{15mm}
          + \frac{3}{2} \xi^3
          \Bigr) \ln^3(\xi) \Biggr] \nn\\
\! \! & & \! \! 
       -  \! \Bigl(
            \frac{5}{2} \xi \! 
          -  \! \frac{21}{4} \xi^2 \! 
          +  \! \frac{17}{2} \xi^3 \! 
          +  \! 3 \xi^4
          \Bigr) \zeta(2) \! 
       +  \!   \Bigl(
            12
          - \frac{149}{5} \xi \! 
          +  \! 28 \xi^2 \! 
          +  \! \frac{39}{5} \xi^3
          - \frac{112}{5} \xi^4
          \Bigr) \zeta^2(2) \nn\\
\! \! & & \! \! 
       +   \Bigl(
            \xi
          - 3 \xi^2
          + \frac{3}{2} \xi^3
          \Bigr) \zeta(3)
       -   \Bigl(
            11
          - \frac{43}{2} \xi
          + \frac{209}{4} \xi^2
          - 57 \xi^3
          + 36 \xi^4
          \Bigr) \zeta(2) \ln(1 - \xi) \nn\\
\! \! & & \! \! 
       -   \Bigl(
            10
          - \frac{23}{2} \xi
          + 9 \xi^2
          - \frac{3}{2} \xi^3
          + 4 \xi^4
          \Bigr) \zeta(3) \ln(1 - \xi)
       -   \Bigl(
            21
          - \frac{73}{2} \xi
          + \frac{119}{4} \xi^2
          - 7 \xi^3 \nn\\
\! \! & & \! \! 
          + \! \xi^4
          \Bigr) \zeta(2) \ln^2(1 - \xi) \! 
       -  \!   \Bigl(
            \frac{1}{2} \! 
          -  \! \frac{29}{24} \xi \! 
          +  \! \frac{73}{48} \xi^2 \! 
          - \!  \frac{5}{4} \xi^3 \! 
          +  \! \frac{7}{12} \xi^4
          \Bigr) \ln^4(1 - \xi) \! 
       -  \!   \Bigl(
            \frac{1}{6} \xi \! 
          -  \! \frac{1}{6} \xi^2 \nn\\
\! \! & & \! \! 
          -  \! \frac{1}{3} \xi^3 \! 
          +  \! \frac{1}{3} \xi^4
          \Bigr) \ln^3(1 - \xi) \! 
       +  \!   \Bigl(
            5 \! 
          -  \! \frac{15}{2} \xi \! 
          +  \! \frac{15}{2} \xi^3 \! 
          -  \! 5 \xi^4
          \Bigr) \ln^2(1 - \xi) {\mathrm{Li}}_{2}(\xi) \! 
       -   \!  \Bigl(
            \frac{1}{2} \! 
          -  \! \frac{15}{4} \xi \nn\\
\! \! & & \! \! 
          + \frac{9}{4} \xi^2
          - \frac{15}{4} \xi^3
          + \frac{1}{2} \xi^4
          \Bigr) \ln^2(1 - \xi)
       +   \Bigl(
            2
          - \frac{5}{2} \xi
          + \frac{5}{2} \xi^3
          - 2 \xi^4
          \Bigr) \ln(1 - \xi) {\mathrm{Li}}_{2}(\xi) \nn\\
\! \! & & \! \! 
       +   \Bigl(
            6
          - 9 \xi
          - \frac{7}{2} \xi^2
          + 14 \xi^3
          - 8 \xi^4
          \Bigr) \ln(1 - \xi) {\mathrm{Li}}_{3}(1 - \xi)
       -   \Bigl(
            2
          - 4 \xi
          + \frac{7}{2} \xi^2 \nn\\
\! \! & & \! \! 
          - \xi^3 \!
          \Bigr) \ln(1 - \xi) {\mathrm{Li}}_{3}\left( - \frac{\xi}{(1-\xi)} \right) \! 
       + \!  \Bigl( \!
            8
          - \frac{17}{2} \xi \! 
          +  \! 4 \xi^2 \! 
          +  \! \frac{3}{2} \xi^3 \! 
          +  \! 2 \xi^4 \!
          \Bigr) \ln(1 - \xi) {\mathrm{Li}}_{3}(\xi) \nn\\
\! \! & & \! \! 
       +  \!  \Bigl( \!
            8
          - 9 \xi \!
          +  \!\frac{29}{2} \xi^2 \!
          - 9 \xi^3 \!
          +  \!8 \xi^4
          \Bigr) \ln(1 - \xi) \!
       +  \!  \Bigl( \!
            12 \xi \! 
          +  \!17 \xi^2 \!
          -  \!26 \xi^3 \!
          +  \!38 \xi^4 \!
          \Bigr) \zeta(2) \ln(\xi) \nn\\
\! \! & & \! \! 
       +   \Bigl(
            6
          - \frac{5}{2} \xi
          + \xi^2
          - \frac{3}{2} \xi^3
          + 4 \xi^4
          \Bigr) \zeta(3) \ln(\xi)
       +   \Bigl(
            40
          - \frac{115}{2} \xi
          + \frac{139}{2} \xi^2
          - \frac{113}{2} \xi^3 \nn\\
\! \! & & \! \! 
          +  \!42 \xi^4
          \Bigr) \zeta(2) \ln(\xi) \ln(1 - \xi) \!
       +   \! \Bigl(
            5
          - \frac{15}{2} \xi \!
          + \! \frac{5}{3} \xi^2 \!
          + \! \frac{9}{2} \xi^3
          - \frac{8}{3} \xi^4
          \Bigr) \ln(\xi) \ln^3(1 - \xi) \nn\\
\! \! & & \! \! 
       +   \Bigl(
            1
          - \frac{3}{4} \xi
          + \frac{1}{8} \xi^2
          + \frac{1}{4} \xi^3
          - \xi^4
          \Bigr) \ln(\xi) \ln^2(1 - \xi)
       -   \Bigl(
            12
          - \frac{33}{2} \xi
          + 9 \xi^2 \nn\\
\! \! & & \! \! 
          + \! \frac{3}{2} \xi^3
          \Bigr) \ln(\xi) \ln(1 - \xi) {\mathrm{Li}}_{2}(\xi)
       -   \Bigl(
            12
          - \frac{43}{4} \xi \! 
          +  \! 10 \xi^2 \! 
          +  \! \frac{3}{2} \xi^3
          - \xi^4
          \Bigr) \ln(\xi) \ln(1 - \xi) \nn\\
\! \! & & \! \! 
       -   \Bigl(
            12
          - \frac{45}{4} \xi
          + 21 \xi^2
          - \frac{101}{4} \xi^3
          + 22 \xi^4
          \Bigr) \zeta(2) \ln^2(\xi)
       -   \Bigl(
            \frac{11}{2}
          - \frac{49}{8} \xi
          + \frac{5}{4} \xi^2 \nn\\
\! \! & & \! \! 
          + \frac{19}{8} \xi^3
          \Bigr) \ln^2(\xi) \ln^2(1 - \xi)
       +   \Bigl(
            \frac{7}{2}
          - 6 \xi
          + \frac{33}{8} \xi^2
          + \frac{1}{4} \xi^3
          - \xi^4
          \Bigr) \ln^2(\xi) \ln(1 - \xi) \nn\\
\! \! & & \! \! 
       +   \Bigl(
            \frac{1}{3}
          + \frac{5}{4} \xi
          - \frac{11}{12} \xi^2
          - \frac{11}{12} \xi^3
          + \frac{5}{3} \xi^4
          \Bigr) \ln^3(\xi) \ln(1 - \xi)
       -   \Bigl(
            \frac{49}{48} \xi
          - \frac{17}{16} \xi^2
          + \frac{1}{12} \xi^3 \nn\\
\! \! & & \! \! 
          + \frac{5}{12} \xi^4
          \Bigr) \ln^4(\xi)
       +   \Bigl(
            \frac{11}{6} \xi
          - \frac{25}{24} \xi^2
          + \frac{5}{24} \xi^3
          + \frac{1}{3} \xi^4
          \Bigr) \ln^3(\xi)
       +   \Bigl(
            6
          - \frac{31}{4} \xi
          + \frac{31}{4} \xi^2 \nn\\
\! \! & & \! \! 
          - \frac{19}{4} \xi^3
          + 3 \xi^4
          \Bigr) \ln^2(\xi) {\mathrm{Li}}_{2}(\xi)
       +   \Bigl(
            \frac{5}{4} \xi
          - \frac{11}{8} \xi^2
          + \frac{1}{4} \xi^3
          - \frac{1}{2} \xi^4
          \Bigr) \ln^2(\xi)
       -   \Bigl(
            4
          - 4 \xi \nn\\
\! \! & & \! \! 
          + \! \frac{3}{2} \xi^2 \! 
          -  \! \frac{3}{2} \xi^3 \! 
          +  \! 2 \xi^4
          \Bigr) \ln(\xi) {\mathrm{Li}}_{2}(\xi) \! 
       -   \!  \Bigl(
            6 \! 
          -  \! 4 \xi \! 
          - \!  \frac{13}{2} \xi^2 \! 
          +  \! 9 \xi^3 \! 
          -  \! 2 \xi^4
          \Bigr) \ln(\xi) {\mathrm{Li}}_{3}(1 - \xi) \nn\\
\! \! & & \! \! 
       +   \Bigl(
            2
          - 4 \xi
          + \frac{7}{2} \xi^2
          - \xi^3
          \Bigr) \ln(\xi) {\mathrm{Li}}_{3}\left( - \frac{\xi}{(1-\xi)} \right)
       -   \Bigl(
            6
          - \frac{3}{2} \xi
          + 2 \xi^2
          - \frac{9}{2} \xi^3 \nn\\
\! \! & & \! \! 
          + 4 \xi^4
          \Bigr) \ln(\xi) {\mathrm{Li}}_{3}(\xi)
       -   \Bigl(
            \xi
          + \frac{29}{4} \xi^2
          - 10 \xi^3
          + 8 \xi^4
          \Bigr) \ln(\xi)
       +   \Bigl(
            22
          - 31 \xi
          + \frac{93}{2} \xi^2 \nn\\
\! \! & & \! \! 
          - 44 \xi^3
          + 32 \xi^4
          \Bigr) \zeta(2) {\mathrm{Li}}_{2}(\xi)
       +   \Bigl(
            \frac{3}{2} \xi
          - \frac{7}{2} \xi^2
          + \frac{7}{2} \xi^3
          - 2 \xi^4
          \Bigr) {\mathrm{Li}}_{3}(1 - \xi)
       -   \Bigl(
            2
          - 4 \xi \nn\\
\! \! & & \! \! 
          + \frac{7}{2} \xi^2
          - \xi^3
          \Bigr) {\mathrm{Li}}_{3}\left( - \frac{\xi}{(1-\xi)} \right)
       +   \Bigl(
            2
          - 2 \xi^2
          - \frac{1}{2} \xi^3
          + 2 \xi^4
          \Bigr) {\mathrm{Li}}_{3}(\xi)
       +   \Bigl(
            7 \xi
          - \frac{9}{2} \xi^2 \nn\\
\! \! & & \! \! 
          - 4 \xi^3 \! 
          +  \! 6 \xi^4
          \Bigr) {\mathrm{Li}}_{4}(1 - \xi) \! 
       -   \!  \Bigl(
            6
          - 4 \xi
          - \frac{9}{2} \xi^2 \! 
          +  \! 7 \xi^3
          \Bigr) {\mathrm{Li}}_{4}\left( - \frac{\xi}{(1-\xi)} \right) \! 
       -   \!  \Bigl(
            2
          - \frac{17}{2} \xi \nn\\
\! \! & & \! \! 
          + \frac{17}{2} \xi^3
          - 2 \xi^4
          \Bigr) {\mathrm{Li}}_{4}(\xi)
          \Biggr\} \, .
\eea

\newpage

\noindent {\bf Including soft radiation} \\

\be
\frac{d \sigma_{2}^{({\tt ph\, 2L \, Box})}}{d \sigma_0} = 
              \delta_{(2)}^{({\tt 2L \, Box},2)} \log^{2}{\left( \frac{s}{m^2} \right)}
            + \delta_{(2)}^{({\tt 2L \, Box},1)} \log{\left( \frac{s}{m^2} \right)} 
	    + \delta_{(2)}^{({\tt 2L \, Box},0)} + {\mathcal O} \left(
	    \frac{m^2}{s} \right) \, ,
\ee	    
where
\bea
\delta_{(2)}^{({\tt 2L \, Box},2)} \! \! \! \! & = & \! \! \! \! \frac{1}{(1-\xi+\xi^2)^2}  (
          - 12
          + 18 \xi
          - 18 \xi^2
          + 6 \xi^3
          ) \, \zeta(2) \, , \\
\delta_{(2)}^{({\tt 2L \, Box},1)} \! \! \! \! & = & \! \! \! \! \frac{1}{(1-\xi+\xi^2)^2} \Bigl\{ 
 \frac{1}{(D-4)} \,   ( 
          - 24
          + 36 \xi
          - 36 \xi^2
          + 12 \xi^3
          ) \, \zeta(2) \nn\\
\! \! \! \! & & \! \! \! \! 
       +  \! \ln^2{\left( \frac{4 \omega^2}{s} \right)}  \Bigl[ 
          (
          - \xi \! 
          + \!  3 \xi^2
          - 3 \xi^3 \! 
          + \!  2 \xi^4
          ) \ln(\xi)
       - (
            2
          - 4 \xi \! 
          + \!  6 \xi^2
          - 4 \xi^3 \! 
          + \!  2 \xi^4
          ) \ln(1 - \xi)  \Bigr] \nn\\
\! \! \! \! & & \! \! \! \! 
       +  \ln{\left( \frac{4 \omega^2}{s} \right)}  \Biggl[ 
          \Bigl( \! 
            2 \xi \! 
          +  \! \frac{3}{2} \xi^2
          - 3 \xi^3 \! 
          +  \! 2 \xi^4 \! 
          \Bigr) \zeta(2)
       -    \Bigl( \! 
            1
          - \frac{5}{2} \xi \! 
          +  \! \frac{7}{2} \xi^2
          - \frac{5}{2} \xi^3 \! 
          +  \! \xi^4 \! 
          \Bigr) \ln^2(1 - \xi) \nn\\
\! \! \! \! & & \! \! \! \! \hspace*{20mm}
       -    \Bigl(
            3
          - \frac{11}{2} \xi
          + 9 \xi^2
          - \frac{11}{2} \xi^3
          + 3 \xi^4
          \Bigr) \ln(1 - \xi)
       -   (
            2
          - 5 \xi
          + 9 \xi^2
          - 7 \xi^3 \nn\\
\! \! \! \! & & \! \! \! \! \hspace*{20mm}
          + 4 \xi^4
          ) {\mathrm Li}_{2}(\xi)
       +   \Bigl(
            2
          - 3 \xi
          + \frac{1}{2} \xi^2
          + 2 \xi^3
          - 2 \xi^4
          \Bigr) \ln(\xi) \ln(1 - \xi) \nn\\
\! \! \! \! & & \! \! \! \! \hspace*{20mm}
       -   \Bigl(
            \xi
          - \frac{9}{2} \xi^2
          + \frac{9}{2} \xi^3
          - 3 \xi^4
          \Bigr) \ln(\xi) 
       +   \Bigl(
            \frac{3}{4} \xi
          - \frac{1}{4} \xi^2
          - \frac{3}{4} \xi^3
          + \xi^4
          \Bigr) \ln^2(\xi)
	  \Biggr] \nn\\
\! \! \! \! & & \! \! \! \! 
       +   \Bigl(
            9 \xi
          - \frac{39}{4} \xi^2
          + \frac{15}{2} \xi^3
          + 3 \xi^4
          \Bigr) \zeta(2)
       -   (
            12 \xi
          - 42 \xi^2
          + 48 \xi^3
          - 24 \xi^4
          ) \zeta(2) \ln(1 - \xi)\nn\\
\! \! \! \! & & \! \! \! \! 
       -   \Bigl(
            \frac{3}{2}
          - \frac{15}{4} \xi
          + \frac{21}{4} \xi^2
          - \frac{15}{4} \xi^3
          + \frac{3}{2} \xi^4
          \Bigr) \ln^2(1 - \xi)
       -   \Bigl(
            \frac{3}{4} \xi
          + \frac{3}{4} \xi^3
          \Bigr) \ln(1 - \xi)\nn\\
\! \! \! \! & & \! \! \! \! 
       -   \Bigl(
            3
          - \frac{15}{2} \xi
          + \frac{27}{2} \xi^2
          - \frac{21}{2} \xi^3
          + 6 \xi^4
          \Bigr) {\mathrm Li}_{2}(\xi)
       -   (
            24
          - 30 \xi
          + 36 \xi^2
          - 30 \xi^3 \nn\\
\! \! \! \! & & \! \! \! \! 
          - 24 \xi^4
          ) \, \zeta(2) \ln(\xi)
       +   \Biggl(
            3
          - \frac{9}{2} \xi
          + \frac{3}{4} \xi^2
          + 3 \xi^3
          - 3 \xi^4
          \Biggr) \ln(\xi) \ln(1 - \xi)
       + \frac{3}{4} \xi \ln(\xi) \nn\\
\! \! \! \! & & \! \! \! \! 
       +   \Biggl(
            \frac{9}{8} \xi
          - \frac{3}{8} \xi^2
          - \frac{9}{8} \xi^3
          + \frac{3}{2} \xi^4
          \Biggr) \ln^2(\xi)
	   \Bigr\} \, , \\
\delta_{(2)}^{({\tt 2L \, Box},0)} \! \! \! \! & = &  \! \! \! \! \frac{1}{(1-\xi+\xi^2)^2} \Biggl\{ \nn\\
\! \! \! \! & & \! \! \! \!  \frac{1}{(D-4)^2}  (
          - 24
          + 36 \xi
          - 36 \xi^2
          + 12 \xi^3
          ) \, \zeta(2) \nn\\
\! \! \! \! & & \! \! \! \!  + \frac{1}{(D-4)} \Bigl[
          (
            6 \xi
          - 12 \xi^2
          + 12 \xi^3
          ) \zeta(2)
       -  (
            12 \xi
          - 42 \xi^2
          + 48 \xi^3
          - 24 \xi^4
          ) \zeta(2) \ln(1 - \xi)\nn\\
\! \! \! \! & & \! \! \! \! \hspace*{15mm}
       -   (
            24
          - 30 \xi
          + 36 \xi^2
          - 30 \xi^3
          + 24 \xi^4
          ) \zeta(2) \ln(\xi) 
	   \Bigr] \nn\\
\! \! \! \! & & \! \! \! \! 
       + \ln^2{\left( \frac{4 \omega^2}{s} \right)}  \Biggl[ 
          (
            1
          - 2 \xi
          + 3 \xi^2
          - 2 \xi^3
          + \xi^4
          ) \ln^2(1 - \xi)
       +   (
            2
          - 4 \xi
          + 6 \xi^2
          - 4 \xi^3 \nn\\
\! \! \! \! & & \! \! \! \! \hspace*{20mm}
          + 2 \xi^4
          ) \ln(1 - \xi)
       -   (
            2
          - 4 \xi
          + 6 \xi^2
          - 4 \xi^3
          + 2 \xi^4
          ) \ln(\xi) \ln(1 - \xi)\nn\\
\! \! \! \! & & \! \! \! \! \hspace*{20mm}
       +   (
            \xi
          - 3 \xi^2
          + 3 \xi^3
          - 2 \xi^4
          ) \ln(\xi)
       -   \Bigl(
            \frac{1}{2} \xi
          - \frac{3}{2} \xi^2
          + \frac{3}{2} \xi^3
          - \xi^4
          \Bigr) \ln^2(\xi) \Biggr] \nn\\
\! \! \! \! & & \! \! \! \! 
       + \ln{\left( \frac{4 \omega^2}{s} \right)}  \Biggl[ 
         \Bigl(
          - 2 \xi
          - \frac{3}{2} \xi^2 \! 
          + \!  3 \xi^3
          - 2 \xi^4
          \Bigr) \zeta(2)
       -    \Bigl(
            \frac{15}{2} \xi^2
          - 9 \xi^3 \! 
          + \!  6 \xi^4
          \Bigr) \zeta(2) \ln(1 - \xi) \nn\\
\! \! \! \! & & \! \! \! \! \hspace*{20mm}
       +    \Bigl(
            1
          - \frac{5}{2} \xi
          + \frac{7}{2} \xi^2
          - \frac{5}{2} \xi^3
          + \xi^4
          \Bigr) \ln^3(1 - \xi)
       +    \Bigl(
            1
          - 2 \xi
          + \frac{7}{2} \xi^2
          - 2 \xi^3 \nn\\
\! \! \! \! & & \! \! \! \! \hspace*{20mm}
          + \xi^4
          \Bigr) \ln^2(1 - \xi)
       +    (
            4
          - 8 \xi
          + 12 \xi^2
          - 8 \xi^3
          + 4 \xi^4
          ) \ln(1 - \xi) {\mathrm Li}_{2}(\xi) \nn\\
\! \! \! \! & & \! \! \! \! \hspace*{20mm}
       +    \Bigl(
            4
          - \frac{15}{2} \xi
          + 12 \xi^2
          - \frac{15}{2} \xi^3
          + 4 \xi^4
          \Bigr) \ln(1 - \xi)
       +    (
            2
          - 5 \xi
          + 9 \xi^2 \nn\\
\! \! \! \! & & \! \! \! \! \hspace*{20mm}
          - 7 \xi^3
          + 4 \xi^4
          ) {\mathrm Li}_{2}(\xi)
       +    \Bigl(
            \frac{15}{2} \xi^2
          - 9 \xi^3
          + 6 \xi^4
          \Bigr) \zeta(2) \ln(\xi)
       -    \Bigl(
            1
          - \frac{5}{2} \xi \nn\\
\! \! \! \! & & \! \! \! \! \hspace*{20mm}
          + \xi^2
          + \frac{1}{2} \xi^3
          - \xi^4
          \Bigr) \ln(\xi) \ln^2(1 - \xi)
       -    \Bigl(
            5
          - \frac{13}{2} \xi
          + 5 \xi^2
          + \xi^3 \nn\\
\! \! \! \! & & \! \! \! \! \hspace*{20mm}
          - 2 \xi^4
          \Bigr) \ln(\xi) \ln(1 - \xi)
       -    (
            2
          - 5 \xi
          + 9 \xi^2
          - 7 \xi^3
          + 4 \xi^4
          ) \ln(\xi) {\mathrm Li}_{2}(\xi) \nn\\
\! \! \! \! & & \! \! \! \! \hspace*{20mm}
       +    \Bigl(
            \frac{3}{2} \xi
          - 6 \xi^2
          + 6 \xi^3
          - 4 \xi^4
          \Bigr) \ln(\xi)
       +    \Bigl(
            2
          - \frac{15}{4} \xi
          + \frac{3}{4} \xi^2
          + \frac{11}{4} \xi^3 \nn\\
\! \! \! \! & & \! \! \! \! \hspace*{20mm}
          - 3 \xi^4
          \Bigr) \ln^2(\xi) \ln(1 - \xi)
       -    \Bigl(
            \frac{1}{4} \xi
          - \frac{1}{4} \xi^2
          - \frac{3}{4} \xi^3
          + \xi^4
          \Bigr) \ln^2(\xi) \nn\\
\! \! \! \! & & \! \! \! \! \hspace*{20mm}
       +    \Bigl(
            \frac{3}{4} \xi
          - \frac{1}{4} \xi^2
          - \frac{3}{4} \xi^3
          + \xi^4
          \Bigr) \ln^3(\xi)
       \Biggr] \nn\\
\! \! \! \! & & \! \! \! \! 
       + \Bigl( 
           \frac{7}{2} \xi
          - 7 \xi^2
          + 4 \xi^3 \Bigr) \zeta(3)
       -   \Bigl(
            \frac{1}{2} \xi
          + \frac{3}{2} \xi^2
          + \xi^3
          + 7 \xi^4
          \Bigr) \zeta(2)
       +   \Bigl(
            12
          - \frac{353}{10} \xi
          + 37 \xi^2 \nn\\
\! \! \! \! & & \! \! \! \! 
          + \frac{3}{10} \xi^3
          - \frac{87}{5} \xi^4
          \Bigr) \zeta^{2}(2)
       -   \Bigl(
            \frac{3}{2} \xi
          + 17 \xi^2
          - \frac{59}{2} \xi^3
          + 18 \xi^4
          \Bigr) \zeta(2) \ln(1 - \xi)
       -   \Bigl(
            13
          - 18 \xi \nn\\
\! \! \! \! & & \! \! \! \! 
          + \frac{19}{2} \xi^2
          + 4 \xi^3
          - 2 \xi^4
          \Bigr) \zeta(2) \ln^{2}(1 - \xi)
       -   \Bigl(
            \frac{1}{4}
          - \frac{1}{8} \xi
          - \frac{11}{48} \xi^2
          + \frac{1}{3} \xi^3
          \Bigr) \ln^{4}(1 - \xi) \nn\\
\! \! \! \! & & \! \! \! \! 
          + \! \frac{2}{3} \xi^2 \ln^{3}(1 - \xi)
       +  \!  \Bigl(
            5
          - \frac{19}{2} \xi \! 
          + \! 7 \xi^2
          - \frac{1}{2} \xi^3
          - \xi^4
          \Bigr) \ln^{2}(1 - \xi) {\mathrm{Li}}_{2}(\xi)
       +  \!  \Bigl(
            \frac{3}{2}
          - \xi \! 
          + 4 \xi^2 \nn\\
\! \! \! \! & & \! \! \! \! 
          - \xi^3
          + \frac{3}{2} \xi^4
          \Bigr) \ln^{2}(1 - \xi)
       +   \Bigl(
            \frac{3}{2} \xi
          + \frac{1}{2} \xi^3
          \Bigr) \ln(1 - \xi) {\mathrm{Li}}_{2}(\xi)
       +   \Bigl(
            6
          - 10 \xi
          + 10 \xi^3 \nn\\
\! \! \! \! & & \! \! \! \! 
          - 6 \xi^4
          \Bigr) \ln(1 - \xi) {\mathrm{Li}}_{3}(1 - \xi)
       + \!   \Bigl(
            6
          - 6 \xi\! 
          + \! \xi^2\! 
          + \! 4 \xi^3
          \Bigr) \ln(1 - \xi) {\mathrm{Li}}_{3}(\xi)
       + \!   \Bigl(
            6 \zeta(3) \xi \! 
          + \! 2 \xi \nn\\
\! \! \! \! & & \! \! \! \! 
          -  \! \zeta(3) \xi^2
          -  \! 4 \zeta(3) \xi^3 \! 
          +  \! 2 \xi^3
          -  \! 6 \zeta(3) \! 
          \Bigr) \ln(1 - \xi) \! 
       +   \!  \Bigl( \! 
            10 \xi \! 
          +  \! \frac{25}{4} \xi^2
          -  \! 6 \xi^3 \! 
          +  \! 18 \xi^4 \! 
          \Bigr) \zeta(2) \ln(\xi) \nn\\
\! \! \! \! & & \! \! \! \! 
       +   \!  \Bigl(
            28
          - 36 \xi \! 
          +  \! \frac{61}{2} \xi^2
          - 23 \xi^3 \! 
          +  \! 22 \xi^4
          \Bigr) \zeta(2) \ln(\xi) \ln(1 - \xi) \! 
       +   \!  \Bigl(
            4
          - \frac{37}{6} \xi \! 
          +  \! \frac{8}{3} \xi^2 \! 
          +  \! \frac{11}{6} \xi^3 \nn\\
\! \! \! \! & & \! \! \! \! 
          - \xi^4 \! 
          \Bigr) \ln(\xi) \ln^{3}(1 - \xi) \! 
       -   \Bigl( \! 
            \frac{3}{2}
          - 4 \xi \! 
          +  \! \frac{29}{8} \xi^2
          - \frac{3}{4} \xi^3 \! 
          \Bigr) \ln(\xi) \ln^{2}(1 - \xi)
       -   \!  \Bigl(
            8
          - 10 \xi \! 
          +  \! \frac{1}{2} \xi^2 \nn\\
\! \! \! \! & & \! \! \! \! 
          +  \! 7 \xi^3
          - 4 \xi^4 \! 
          \Bigr) \ln(\xi) \ln(1 - \xi) {\mathrm{Li}}_{2}(\xi)
       -   \!  \Bigl( \! 
            4
          - \frac{5}{4} \xi
          - 2 \xi^2 \! 
          +  \! 8 \xi^3
          - 5 \xi^4 \! 
          \Bigr) \ln(\xi) \ln(1 - \xi) \nn\\
\! \! \! \! & & \! \! \! \! 
       -   \Bigl(
            12
          - \frac{27}{2} \xi
          + 14 \xi^2
          - \frac{25}{2} \xi^3
          + 11 \xi^4
          \Bigr) \zeta(2) \ln^{2}(\xi)
       -   \Bigl(
            \frac{11}{2}
          - \frac{27}{4} \xi
          + \frac{15}{8} \xi^2
          + 2 \xi^3 \nn\\
\! \! \! \! & & \! \! \! \! 
          - \frac{1}{2} \xi^4
          \Bigr) \ln^{2}(\xi) \ln^{2}(1 - \xi)
       +  \! \Bigl( \! 
            3
          - \frac{15}{2} \xi \! 
          +  \! \frac{29}{4} \xi^2
          - 2 \xi^3 \! 
          \Bigr) \ln^{2}(\xi) \ln(1 - \xi)
       +   \!  \Bigl( \! 
            \frac{2}{3}
          - \frac{17}{12} \xi \nn\\
\! \! \! \! & & \! \! \! \! 
          + \frac{3}{4} \xi^2
          + \frac{5}{12} \xi^3
          - \frac{2}{3} \xi^4
          \Bigr) \ln^{3}(\xi) \ln(1 - \xi)
       +   \Bigl(
            \frac{7}{48} \xi
          - \frac{5}{48} \xi^2
          - \frac{1}{12} \xi^3
          + \frac{1}{6} \xi^4
          \Bigr) \ln^{4}(\xi) \nn\\
\! \! \! \! & & \! \! \! \! 
       +   \Bigl(
            \frac{5}{3} \xi
          - \frac{3}{2} \xi^2
          + \frac{5}{24} \xi^3
          \Bigr) \ln^{3}(\xi)
       +   \Bigl(
            2
          - \frac{11}{4} \xi
          + \frac{7}{4} \xi^2
          + \frac{1}{4} \xi^3
          - \xi^4
          \Bigr) \ln^{2}(\xi) {\mathrm{Li}}_{2}(\xi) \nn\\
\! \! \! \! & & \! \! \! \! 
       +  \! \Bigl(
            \xi
          - 3 \xi^2 \! 
          +  \! \frac{13}{4} \xi^3
          - \frac{5}{2} \xi^4
          \Bigr) \ln^{2}(\xi)
       -   \Bigl(
            3
          - \xi
          - \frac{1}{2} \xi^2 \! 
          +  \! \frac{1}{2} \xi^3
          \Bigr) \ln(\xi) {\mathrm{Li}}_{2}(\xi)
       -   (
            6
          - 5 \xi \nn\\
\! \! \! \! & & \! \! \! \! 
          -  \! 3 \xi^2 \! 
          +  \! 5 \xi^3 \! 
          ) \ln(\xi) {\mathrm{Li}}_{3}(1 - \xi) \! 
       -   \!  (
            4 \! 
          +  \! \xi \! 
          -  \! \xi^2 \! 
          -  \! 2 \xi^3 \! 
          +  \! 2 \xi^4
          ) \ln(\xi) {\mathrm{Li}}_{3}(\xi) \! 
       -   \!  ( 
            3 \zeta(3) \xi \! 
          +  \! 2 \xi \nn\\
\! \! \! \! & & \! \! \! \! 
          +  \! \zeta(3) \xi^2
          - 2 \zeta(3) \xi^3
          - 6 \zeta(3)
          ) \ln(\xi) \! 
       +   \!  \Bigl(
            22
          - 31 \xi \! 
          +  \! \frac{63}{2} \xi^2
          - 26 \xi^3 \! 
          +  \! 20 \xi^4
          \Bigr) \zeta(2) {\mathrm{Li}}_{2}(\xi)  \nn\\
\! \! \! \! & & \! \! \! \! 
       +   \!   \Bigl(
            3
          - \frac{13}{2} \xi \! 
          +  \! \frac{21}{2} \xi^2
          - \frac{15}{2} \xi^3 \! 
          +  \! 4 \xi^4
          \Bigr){\mathrm{Li}}_{2}^{2}(\xi)
       +   \!   \Bigl(
            4
          - 10 \xi
          + 18 \xi^2
          - 14 \xi^3
          + 8 \xi^4
          \Bigr){\mathrm{Li}}_{2}(\xi)  \nn\\
\! \! \! \! & & \! \! \! \! 
       +    \Bigl(
            \frac{1}{2} \xi
          - \frac{1}{2} \xi^3
          \Bigr){\mathrm{Li}}_{3}(1 - \xi)
       +    \Bigl(
            \frac{5}{2} \xi
          - 5 \xi^2
          + 2 \xi^3
          \Bigr){\mathrm{Li}}_{3}(\xi)
       +    \Bigl(
            7 \xi
          - \frac{9}{2} \xi^2
          - 4 \xi^3  \nn\\
\! \! \! \! & & \! \! \! \! 
          + 6 \xi^4
          \Bigr){\mathrm{Li}}_{4}(1 - \xi)
       -    \Bigl(
            6
          - 4 \xi
          - \frac{9}{2} \xi^2
          + 7 \xi^3
          \Bigr){\mathrm{Li}}_{4} \left( - \frac{\xi}{(1 - \xi)} \right)
       -    \Bigl(
            2
          - \frac{17}{2} \xi  \nn\\
\! \! \! \! & & \! \! \! \! 
          + \frac{17}{2} \xi^3
          - 2 \xi^4
          \Bigr){\mathrm{Li}}_{4}(\xi)
	  \Biggr\} \, .
\eea

\newpage


\begin{thebibliography}{99}
\parskip 0pt
\itemsep=0pt


\bibitem{reviews}
  S.~Jadach {\it et al.},
  ``Event Generators for Bhabha Scattering,''
  arXiv:hep-ph/9602393. \\
  G.~Montagna, O.~Nicrosini and F.~Piccinini,
  Riv.\ Nuovo Cim.\  {\bf 21N9} (1998) 1
  [arXiv:hep-ph/9802302].


\bibitem{Bhabha1loop}  
  M.~Consoli,
  Nucl.\ Phys.\ B {\bf 160} (1979) 208.\\
  M.~Bohm, A.~Denner and W.~Hollik, 
  Nucl.\ Phys.\ B {\bf 304} (1988) 687.


\bibitem{russians}
  G.~Faldt and P.~Osland,
  Nucl.\ Phys.\ B {\bf 413} (1994) 64
  [arXiv:hep-ph/9304301]. \\
  G.~Faldt and P.~Osland,
  Nucl.\ Phys.\ B {\bf 413} (1994) 16
  [Erratum-ibid.\ B {\bf 419} (1994) 404]
  [arXiv:hep-ph/9304212]. \\
  A.~B.~Arbuzov, E.~A.~Kuraev and B.~G.~Shaikhatdenov,
  Mod.\ Phys.\ Lett.\ A {\bf 13} (1998) 2305
  [arXiv:hep-ph/9806215]. \\
  A.~B.~Arbuzov, E.~A.~Kuraev, N.~P.~Merenkov and L.~Trentadue,
  Nucl.\ Phys.\ B {\bf 474} (1996) 271.


\bibitem{Fadin:1993ha}
V.~S.~Fadin, E.~A.~Kuraev, L.~Trentadue, L.~N.~Lipatov and N.~P.~Merenkov,
Phys.\ Atom.\ Nucl.\  {\bf 56} (1993) 1537
[Yad.\ Fiz.\  {\bf 56N11} (1993) 145].

\bibitem{Arbuzov:1995vj}
A.~B.~Arbuzov, E.~A.~Kuraev, N.~P.~Merenkov and L.~Trentadue,
Phys.\ Atom.\ Nucl.\  {\bf 60} (1997) 591
[Yad.\ Fiz.\  {\bf 60N4} (1997) 673].

\bibitem{Bas}
E.~W.~N.~Glover, J.~B.~Tausk and J.~J.~Van der Bij,
Phys.\ Lett.\ B {\bf 516} (2001) 33
[arXiv:hep-ph/0106052].

\bibitem{Bern}
  Z.~Bern, L.~J.~Dixon and A.~Ghinculov,
  Phys.\ Rev.\ D {\bf 63} (2001) 053007
  [arXiv:hep-ph/0010075].

\bibitem{bmr}
R.~Barbieri, J.~A.~Mignaco and E.~Remiddi,
Nuovo Cim.\ A {\bf 11} (1972) 824;
Nuovo Cim.\ A {\bf 11} (1972) 865.

\bibitem{Catani}
  S.~Catani,
  Phys.\ Lett.\ B {\bf 427} (1998) 161
  [arXiv:hep-ph/9802439].


\bibitem{Penin:2005kf}
A.~A.~Penin,
arXiv:hep-ph/0501120v3.


\bibitem{us}
R.~Bonciani, A.~Ferroglia, P.~Mastrolia, E.~Remiddi and J.~J.~van der Bij,
Nucl.\ Phys.\ B {\bf 681} (2004) 261
[Erratum-ibid.\ B {\bf 702} (2004) 364]
[arXiv:hep-ph/0310333].

\bibitem{us2}
R.~Bonciani, A.~Ferroglia, P.~Mastrolia, E.~Remiddi and J.~J.~van der Bij,
Nucl.\ Phys.\ B {\bf 701} (2004) 121
[arXiv:hep-ph/0405275].

\bibitem{us3}
  R.~Bonciani, A.~Ferroglia, P.~Mastrolia, E.~Remiddi and J.~J.~van der Bij,
  Nucl.\ Phys.\ B {\bf 716} (2005) 280
  [arXiv:hep-ph/0411321].

%

\bibitem{Lap}
S.~Laporta and E.~Remiddi,
Phys.\ Lett.\ B {\bf 379} (1996) 283
[arXiv:hep-ph/9602417]. \\
S.~Laporta,
Int.\ J.\ Mod.\ Phys.\ A {\bf 15} (2000) 5087
[arXiv:hep-ph/0102033].

%

\bibitem{IBP}
F.~V.~Tkachov,
Phys.\ Lett.\ B {\bf 100} (1981) 65.\\
G.~Chetyrkin and F.~V.~Tkachov,
Nucl.\ Phys.\ B {\bf 192} (1981) 159.

\bibitem{LI}
T.~Gehrmann and E.~Remiddi,
Nucl.\ Phys.\ B {\bf 580} (2000) 485
[arXiv:hep-ph/9912329].


\bibitem{DiffEq}
A.~V.~Kotikov,
Phys.\ Lett.\ B {\bf 254} (1991) 158. \\
A.~V.~Kotikov,
Phys.\ Lett.\ B {\bf 259} (1991) 314. \\
A.~V.~Kotikov,
Phys.\ Lett.\ B {\bf 267} (1991) 123. \\
E.~Remiddi,
Nuovo Cim.\ A {\bf 110} (1997) 1435
[arXiv:hep-th/9711188]. \\
M.~Caffo, H.~Czyz, S.~Laporta and E.~Remiddi,
Acta Phys.\ Polon.\ B {\bf 29} (1998) 2627
[arXiv:hep-th/9807119]. \\
M.~Caffo, H.~Czyz, S.~Laporta and E.~Remiddi,
Nuovo Cim.\ A {\bf 111} (1998) 365
[arXiv:hep-th/9805118].




\bibitem{HPLs}
  A.B.Goncharov, 
  Math. Res. Lett. 5 (1998), 497-516.\\  
  D.~J.~Broadhurst,
  Eur.\ Phys.\ J.\ C {\bf 8} (1999) 311
  [arXiv:hep-th/9803091]. \\
  E.~Remiddi and J.~A.~M.~Vermaseren,
  Int.\ J.\ Mod.\ Phys.\ A {\bf 15} (2000) 725
  [arXiv:hep-ph/9905237].\\
  T.~Gehrmann and E.~Remiddi,
  Comput.\ Phys.\ Commun.\  {\bf 141} (2001) 296
  [arXiv:hep-ph/0107173]. \\
  T.~Gehrmann and E.~Remiddi,
  Nucl.\ Phys.\ B {\bf 601} (2001) 248
  [arXiv:hep-ph/0008287]. \\
  T.~Gehrmann and E.~Remiddi,
  Nucl.\ Phys.\ B {\bf 640} (2002) 379
  [arXiv:hep-ph/0207020]. \\
  T.~Gehrmann and E.~Remiddi,
  Comput.\ Phys.\ Commun.\  {\bf 144} (2002) 200
  [arXiv:hep-ph/0111255].


\bibitem{DimReg}
  G.~'t Hooft and M.~J.~G.~Veltman,
  Nucl.\ Phys.\ B {\bf 44} (1972) 189.\\
  C. G. Bollini and J. J. Giambiagi, 
  Phys.\ Lett.\ B {\bf 40} (1972) 566;
  Nuovo Cim.\ B {\bf 12} (1972) 20. \\
  J.~F.~Ashmore,
  Lett.\ Nuovo Cim.\  {\bf 4} (1972) 289. \\
  G.~M.~Cicuta and E.~Montaldi,
  Lett.\ Nuovo Cim.\  {\bf 4} (1972) 329. \\
  R.~Gastmans and R.~Meuldermans,
  Nucl.\ Phys.\ B {\bf 63} (1973) 277.




\bibitem{Czakon:2004tg}
M.~Czakon, J.~Gluza and T.~Riemann,
Nucl.\ Phys.\ Proc.\ Suppl.\  {\bf 135} (2004) 83
[arXiv:hep-ph/0406203].

\bibitem{Czakon:2004wm}
  M.~Czakon, J.~Gluza and T.~Riemann,
  Phys.\ Rev.\ D {\bf 71} (2005) 073009
  [arXiv:hep-ph/0412164].


\bibitem{Gundrum}
  V.~A.~Smirnov,
  Phys.\ Lett.\ B {\bf 524} (2002) 129
  [arXiv:hep-ph/0111160].\\
  G.~Heinrich and V.~A.~Smirnov,
  Phys.\ Lett.\ B {\bf 598} (2004) 55
  [arXiv:hep-ph/0406053].

%
\bibitem{Fleischer:2002wa}
  J.~Fleischer, T.~Riemann, O.~V.~Tarasov and A.~Werthenbach,
  Nucl.\ Phys.\ Proc.\ Suppl.\  {\bf 116} (2003) 43
  [arXiv:hep-ph/0211167].



\bibitem{RoPieRem1}  
  R.~Bonciani, P.~Mastrolia and E.~Remiddi,
  Nucl.\ Phys.\ B {\bf 661} (2003) 289
  [Erratum-ibid.\ B {\bf 702} (2004) 359]
  [arXiv:hep-ph/0301170].

\bibitem{RoPieRem2}   
  R.~Bonciani, P.~Mastrolia and E.~Remiddi,
  Nucl.\ Phys.\ B {\bf 676} (2004) 399
  [arXiv:hep-ph/0307295].

%

\bibitem{file}
The expression of the functions introduced in Sections~\ref{sec1}--\ref{secN} are
included in the file {\tt Vfactors.inc}; the expression of the functions 
introduced in Section~\ref{sec4} are included in the file {\tt Bfactors.inc}.
The expansion in the limit $m^2/s \to 0$ of all the contributions to the 
differential cross section discussed in the paper are collected in 
{\tt Expansions.inc}. The files can be downloaded from 
{\tt pheno.physik.uni-freiburg.de/\~\,bhabha}.



\bibitem{GP}
 D.~Bardin and G.~Passarino ,
``{\it The standard model in the making: Precision study of the electroweak
interactions}'', Oxford University Press, 1999.


\bibitem{FORM}
  J.A.M. Vermaseren, {\it Symbolic Manipulation with}
  {\tt FORM}, Version 2, CAN, Amsterdam, 1991; 
  New features of {\tt FORM}, ({\tt math-ph/0010025}).


\end{thebibliography}
\end{document}